\documentclass[twocolumn,tighten,times]{aastex62}
\pdfoutput=1

\received{July 26, 2018}
\revised{February 10, 2019}
\accepted{February 13, 2019}
\submitjournal{ApJ}

\begin{document}
\title{Timing Properties of Shocked Accretion Flows around Neutron Stars in Presence of Cooling}

\correspondingauthor{Ayan Bhattacharjee}
\email{ayan12@bose.res.in, chakraba@bose.res.in}
\author[0000-0002-2878-4025]{Ayan Bhattacharjee,}
\affiliation{S. N. Bose National Centre for Basic Sciences, Salt Lake, Kolkata 700106, India}
\author[0000-0002-0193-1136]{Sandip K. Chakrabarti}
\affiliation{S. N. Bose National Centre for Basic Sciences, Salt Lake, Kolkata 700106, India}
\affiliation{Indian Center for Space Physics, 43 Chalantika, Garia St. Road, Kolkata 700084, India}

\label{firstpage}
\begin{abstract}
We carry out the first robust numerical simulation of accretion flows on a weakly magnetized neutron star using Smoothed Particle Hydrodynamics (SPH). We follow the Two-Component Advective Flow (TCAF) paradigm for black holes, and focus only on the advective component for the case of a neutron star. This low viscosity sub-Keplerian flow will create a normal boundary layer (or, NBOL) right on the star surface in addition to the centrifugal pressure supported boundary layer (or, CENBOL) present in a black hole accretion. These density jumps could give rise to standing or oscillating shock fronts. During a hard spectral state, the incoming flow has a negligible viscosity causing more sub-Keplerian component as compared to the Keplerian disc component. We show that our simulation of flows with a cooling and a negligible viscosity produces precisely two shocks and strong supersonic winds from these boundary layers. We find that the specific angular momentum of matter dictates the locations and the nature of oscillations of these shocks. For low angular momentum flows, the radial oscillation appears to be preferred. For flows with higher angular momentum, the vertical oscillation appears to become dominant. In all the cases, asymmetries w.r.t. the Z=0 plane are seen and instabilities set in due to the interaction of inflow and outgoing strong winds. Our results capture both the low and high-frequency quasi-periodic oscillations without invoking magnetic fields or any precession mechanism. Most importantly, these solutions directly corroborate observed features of wind dominated high-mass X-ray binaries, such as Cir X-1.
\end{abstract}


\section{Introduction}
Modelling of accretion process around neutron stars is largely prompted by significant observational findings ever since they were discovered. Although many of the models are of great phenomenological importance, a unified solution is still not present that describes both the spectral and temporal behavior self-consistently. For black holes, a self-consistent solution that addresses timing and spectral properties simultaneously, exists in the form of Two-Component Advective Flow (hereafter TCAF; Chakrabarti, 1995, 1997, see also, Chakrabarti \& Titarchuk 1995 for some spectra of TCAF) where both the Keplerian and sub-Keplerian (low angular momentum) matter accrete together. With a goal to study the applicability TCAF paradigm for neutron stars, we conduct a 2D hydrodynamic simulation of inviscid sub-Keplerian accreting flow around a non-magnetic neutron star. We argue that when the accretion rate is low and viscosity parameter $\alpha$ is negligible, our simulations show shocks. When the viscosity is higher, a two-component advective flow (TCAF) would disaggregate out of a single sub-Keplerian component flow, as happened, for example in black hole accretion (Chakrabarti, 1990; Chakrabarti, 1996, Giri \& Chakrabarti, 2013). 

The temporal variations of both magnetic and weakly-magnetic, accreting NSs are reflected in the Power Density Spectra (PDS) of the lightcurves at different energies. The presence and evolution of QPOs reveal significant details about both the hydrodynamic and radiative transfer processes. We have used the definitions from Wang (2016) for the different classes of Quasi-Periodic Oscillations (QPOs) and the corresponding abbreviations have been used throughout this paper.
\begin{itemize}
\item \textbf{Low-Frequency QPOs (LFQPO):} Low-frequency QPOs, with Quality Factor $Q \geq 2$, and amplitudes of $1\%-10\%$, are observed in the range of 5 - 60 Hz. The Flaring-, Normal-, and Horizontal Branch Oscillations all lie in this domain and are respectively abbreviated as FBO, NBO and HBO.
\item \textbf{hecto-Hz QPOs (hHz QPO):} A peaked noise usually shows up as the hecto-hertz (hHz) QPO. If the peak is coherent enough to have a quality factor $Q>2$, it is classified as a QPO. The frequency $\nu_{hHz}$ is seen in the range of $100-200$ Hz. It has rms amplitude between $2\%$ and $20\%$.
\item \textbf{Kilo Hz QPOs (kHz QPO):} These are, usually, defined as the QPOs in the range $200~Hz < \nu_{kHz} < 1300~Hz$.
\end{itemize}

We take the following approach to address the validity of our solution: we point to the observational evidences that either supported the TCAF paradigm, or refuted an alternate theory in light of new results. We have touched upon the relevant models in the literature and compared them in terms of predictions of observational results, as an introduction to the studies reported here by us. A more detailed version of this, which is summarized here, can be found in Bhattacharjee (2018).

Elsner \& Lamb (1977) carried out the first detailed study of almost radial accretion onto a magnetic neutron star where discussed important properties of the magnetosphere. The spin-up and spin-down mechanism of the star due to this was studied by Ghosh, Lamb \& Pethick (1977). The structures of the transition region and the changing of period of the pulsating stars were studied by Ghosh \& Lamb (1979a, 1979b). The study of the power density spectra of the source GX 5-1 by Van der Klis et al. (1985) revealed showed intensity dependence of the centroid frequency, width and power of the observed QPOs (20 Hz and 40 Hz) and low-frequency noises ($< 15~Hz$). The interaction of a clumpy disc and a weak magnetosphere was proposed as the cause of the modulation in accretion rate and oscillations in X-ray flux of the source Sco X-1 which reflected as QPOs in 5-50 Hz range. A bimodal behavior in the response of QPO frequency with intensity was found for Sco X-1 by Priedhorsky et al. (1986), where they showed that the 6 Hz QPO, in quiescent state, was anti-correlated with intensity, whereas in the active state the 10-20 Hz QPOs were correlated with intensity. An unsteady flow with low viscosity was proposed to be the cause of the luminosity variations of the boundary by Paczynski (1987). Correlation between the spectral shape, fast variability, temperature and burst duration with accretion rate $\dot{M}$ as well as lack thereof with persistent intensity for 4U 1636-53 lead Van der Klis et al. (1990) to suggest that the accretion rate $\dot{M}$, which might not be well measured by intensity, determines the source state. Strohmayer et al. (1996) used magnetospheric beat frequency model to explain the constant separation of 363 Hz, which was equal to the burst pulsations, between the twin kHz QPOs that varied between 650-1100 Hz, for the LMXB 4U 1728-34. For Sco X-1, Van der Klis et al. (1997) found: 1) the twin kHz QPO frequency separation, rather than being constant, varied from 310 to 230 Hz and concluded that these peaks are not likely to be explained by any photon bubble model or beat frequency model; 2) both the HBOs, near 45 Hz and 90 Hz, increased with accretion rate. The energy dependent study of twin kHz QPOs and band limited noise by M{\'e}ndez et al. (1997) suggested that the kHz QPOs were due to oscillations of the boundary layer, very close to the surface and the broad noise was generated from near the inner edge of a disc and these frequencies tracked $\dot{M}$ better than the count rate. Simultaneous HBOs and kHz QPOs for GX 17+2, found by Wijnands et al. (1997), suggested that the phenomena could not be simultaneously explained by the magnetospheric beat frequency model. The rms and FWHM of the lower kHz QPO remained constant with changes in frequency, but that of upper kHz QPO varied. The Z-source Cyg X-2 also discovered to exhibit simultaneous kHz QPOs and HBOs by Wijnands et al. (1998). For the Z-source GX 340+0, Jonker et al. (1998) showed that the twin kHz QPO frequencies moved to higher values with the increase of accretion rate. The rms and FWHM of the upper kHz QPO decreased, whereas the ones corresponding to the lower QPO remained generally constant. Simultaneous HBO was also detected along with its second harmonic between 20 to 50 Hz and 38 Hz and 69 Hz. They also concluded that something other than $\dot{M}$ determined the timing properties, from the study of FWHM of HBOs with states. Wijnands et al. (1998) observed simultaneous kHz QPOs and HBOs for the object GX 5-1 and found similar results.

In Titarchuk et al. (1998, hereafter TLM98) a super-Keplerian transition layer (TL) was invoked to explain the kHz QPOs. M{\'e}ndez et al. (1998) showed a varying separation between the centroid frequencies of twin kHz QPOs of LMXB 4U 1608-522 and ruled out any simple beat-frequency model. Simultaneous existence of burst oscillation and twin KHz QPOs for 4U 1728-34 was found by M{\'e}ndez \& van der Klis (1999). For the same object, the TL model was used by Titarchuk \& Osherovich (1999) where they attributed radial oscillation in viscous time-scale and radial diffusion time-scale as the possible determining factors behind low-Lorentzian frequency and the break frequency, respectively. A correlation between 1)the NS: kHz QPOs and HBOs, 2)BH: QPOs and noises, of same type but varying 3 orders of magnitude in frequency and coherence was found by Psaltis et al. (1999) which suggested that the variations are systematic and related by similar processes in the two types of compact objects. Different modes of QPOs observed for 4U 1728-34, were explained by adding the effects of Coriolis forces and interaction of magnetosphere by Titarchuk et al. (1999). The study of Sco X-1 by Dieters et al. (2000) revealed that the position on the Z-track or the spectral state was not the only parameter that governs the behavior of the source. A new broad component in the PDS of GX 340+0 between 9 Hz to 14 Hz was discovered by Jonker et al. (2000). For the source Sco X-1, Yu et al. (2001) the ratio of lower to upper kHz QPO amplitude and the upper kHz QPO frequency were anti-correlated with the count rate that varied in the NBO timescale (6-8 Hz) and suggested some of the NBO flux was generated from inside the inner disc radius and the radiative stress modulated the NBO frequency. A long term study of a total of 13 Z and atoll sources by Muno et al. (2002) showed that similar three branched color-color patterns are traced out by both, which was previously not recorded because of incomplete sampling, suggested a similarity between the two types of sources.

The decomposition of PDS into 4 primary Lorentzians for both BHs and NSs by Belloni et al. in 2002 revealed that the physical processes of HFQPOs might be the same for both types of compact object, and possibly independent of the stellar surface. In the same year, Mauche extended the QPO correlation between BH, NS to the case of White Dwarfs, over 5 orders of magnitudes, further suggesting a common mechanism of oscillations in these systems. From their studies of multiple accreting NSs, Wijnands et al. (2003) suggested that the QPOs can be understood in terms of resonances at specific radii in an accretion disc. Barret et al. (2005) ruled out the possibility of any model involving clumps and disc/shock oscillations were suggested to be the more likely mechanism of generating QPOs in LMXB 4U 1608-52. Belloni et al. (2005) showed that a biased sampling leads to the finding of ratio of upper and lower kHz QPO frequencies around 3/2, which indicated that a simple resonance model of a Keplerian disc is unlikely to address the kHz QPOs. From the one-to-one correspondence between the LFQPOs in BHs and NSs, Cassella et al. (2006) inferred that the physical mechanisms that determine these oscillations, are in favor of a disc origin of oscillation and rules out models involving interaction with the magnetosphere or the surface. M{\'e}ndez (2006) suggested that one possible mechanism of generation of kHz QPOs would be that the mass accretion rate sets the size of the inner radius of the disc which determines the QPO frequency as well as the relative contribution of the high-energy part of the spectrum to the total luminosity. To explain the lower value of kHz QPOs in Cir X-1, Boutloukos et al. (2006) pointed that a high radially accreting flow is more likely than a Keplerian disc for the source. The study of NBOs for Sco X-1 by Wang et al. 2012 also suggested a radial oscillation as the centroid frequency varied non-monotonically with energy.

\section{TCAF around an NS}
\label{sec:2}

In Bhattacharjee and Chakrabarti (2017), we discussed at length how the spectral signature of TCAF are present in observations of accretion flows around neutron stars. Here, we include more of those features which pertain to the timing properties of such flows and how there is a connection between the TCAF scenario and the observed results. A detailed version of this comparative study is reported in Bhattacharjee (2018), which we summarize here. The spectral and timing properties of accreting matter around a black hole could not be explained by a standard Keplerian disc (Sunyaev \& Truemper, 1979, Haardt and Maraschi 1993, Zdziarski et al. 2003; Chakrabarti 1995; CT95; Chakrabarti, 1997). The two main components of the spectrum are: 1. a thermal component which resembles a multi-colour blackbody radiation (Shakura \& Sunyaev 1973), 2. a power-law-like component generated by inverse Comptonization of the thermal or non-thermal electrons (Sunyaev \& Titarchuk, 1980, 1985). Although many models deal with the production of a Compton cloud that generates the power-law, many such possible scenarios required excessive fine-tuning of accreting matter and are self-consistent or not physical (for a review, see Chakrabarti 2017). TCAF, which is based on the transonic flow solution, is a unique self-consistent solution (Chakrabarti, 1995, 1996, 1997), equally applicable for BHs and NSs when proper inner boundary condition is chosen. All the aspects of spectral and temporal properties are addressed at the same time by the TCAF paradigm. In this scenario, the low viscosity advection component produces a centrifugal barrier supported boundary layer or CENBOL very close to a compact object. A shock transition defines the boundary of CENBOL. This shock surface may be stationary, oscillating or propagatory (Chakrabarti 2017 and references therein). The post-shock region behaves as a natural reservoir of hot electrons. The highly viscous component of the flow near the equatorial plane becomes a Keplerian disc (Giri \& Chakrabarti 2013). This disc emits a multi-colour blackbody radiation which are intercepted and re-radiated by the CENBOL to create the power-law-like component (Chakrabarti, 1997; Garain et al. 2014, hereafter GGC14). The power-law component usually has an exponential cut-off where recoil becomes important.

The shock formed in the transonic flows around BHs and NSs can oscillate due to resonance (Molteni et al. 1996, hereafter MSC96) or non-satisfaction of Rankine-Hugoniot condition (Ryu et al. 1997). The variation in the size of the CENBOL results in modulation of intensity of the hard X-ray, manifesting as the LFQPOs. The CENBOL also acts as the source of outflows and jets. When excessive soft photons cool the CENBOL down, the jet disappears as well. There are clear observational evidence of two-component advective flows in several black hole candidates (Smith et al. 2001; Smith et al. 2002; Debnath et al. 2013; Mondal et al. 2014; Dutta \& Chakrabarti, 2016; Bhattacharjee et al. 2017; Ghosh and Chakrabarti, 2018).

A natural explanation of the phenomenological Compton clouds for the black hole accretion can be found in the TCAF paradigm. Apart from the region near the innermost boundary, the flow accreting from the outer edge of the disc has little knowledge about the nature of the compact object. Therefore, we anticipate, especially when the magnetic field is weak ($<10^8$~Gauss), even an NS accretion would have a CENBOL which forms away from the surface of the NS. The advective matter almost freely falls under gravity till it reaches the surface of the NS. Thus, in a neutron star accretion, two shocked layers woould be simultaneously present: One is similar to the normal boundary layer (NBOL), and the other is similar to the CENBOL in a black hole accretion (Chakrabarti 1995). Only CENBOL is present in a black hole accretion. In the following segment, we point to some of the key similarities between the TCAF scenario and the flow configuration demanded by observations. The conclusions of Chakrabarti (1996) and Chakrabarti \& Sahu (1997) that the boundary condition of the gravitating object (BH or NS) modifies the solutions of the transonic flows only in the last few Schwarzschild radii, are strengthened by these points.

The following observed features can be reproduced when the accretion flow on a neutron star is TCAF:
\begin{itemize}
\item The Low-Frequency QPOs ($\nu_{LF}$) shows a bimodal behaviour (Priedhorsky et al. 1986): The intensity is controlled by two separate accretion rates. One increases the oscillation radius, another reduces it.
\item BHs, NSs and WDs show correlated QPOs (Psaltis et al. 1999, Belloni et al. 2002, Mauche 2002): A stellar surface or magnetosphere should not be required by the generalized model.
\item Sources of varying intensity having similar kHz QPO frequency ($\nu_{kHz}$) (M{\'e}ndez et al. 1998): More than one accretion rates control the QPOs.
\item Long coherence time of kHz QPOs (Barret and Olive 2005): Vertical and radial oscillations would be preferred. It is unlikely that clumpy disc models with azimuthal asymmetry would generate QPOs in Keplerian orbital time-scale.
\item A `mass' accretion rate which controls the relative contribution to high energy part of the spectrum also decides the inner edge of the disc that generates kHz QPOs (M{\'e}ndez 2006).
\item The Compton cloud acts as the base of the jet (Paizis. et al. 2006).
\item The state transition was controlled by more than just a single accretion rate (Barret 2001; Barret and Olive 2002).
\item Some NSs (For Cir X-1, Boutloukos et al. 2006) show low $\nu_{kHz}$: An advective flow (radial accretion) is preferred over a Keplerian disc.
\end{itemize}

We divide our study of accretion flow into two broad classes: 1. The flow is inviscid, advective and has a low efficiency of radiation. It also comprises of winds from the companion star. 2. The flow has a significant viscosity to create two-component advective flow and has a blackbody radiation emitting disc. 

The main focus of the present work is to concentrate on  Class 1 flows, namely, the effects of angular momentum of a sub-Keplerian flow in absence of significant dynamic viscosity. In such a scenario, which can occur when the accretion rate is relatively low as compared to the Eddington limit, we assume the bremsstrahlung cooling to be the dominant process. These classes are not that well explored in the literature as in the case of black hole
accretion using hydrodynamic simulations, and thus we focus on this domain of solutions first.
However, when the viscosity is high enough to produce a Keplerian disc on the equatorial plane, the blackbody emission from the disc as well as the non-local cooling processes such as Comptonization would become significant. Those aspects of Class 2 flows are not within the scope of this paper and will be discussed in a subsequent paper (Bhattacharjee and Chakrabarti, in preparation). TLM98 and subsequent works have tried to explore this domain, although with the assumption that the entire disc is Keplerian to begin with. In Bhattacharjee \& Chakrabarti (2017, hereafter BC17), we have shown similarities and differences between the TLM98 scenario and TCAF paradigm in spectral analysis and will carry out timing study in a subsequent paper.

\section{Method}
The Smoothed Particle Hydrodynamics (SPH) method was introduced by Monaghan (1992). 
It has since been used in many astrophysical systems, including simulation of accreting matter around black holes to simulate accretion in 1D (Chakrabarti \& Molteni 1993, hereafter CM93); 2D flows (Molteni, Lanzafame \& Chakrabarti 1994, hereafter MLC94); viscous Keplerian discs (Chakrabarti \& Molteni 1995, hereafter CM95); resonance oscillation of shocks due to cooling in 2D (MSC96); comparative study of shocked advective flows using SPH and TVD schemes (Molteni, Ryu \& Chakrabarti 1996, hereafter MRC96); thick accretion discs (Lanzafame, Molteni \& Chakrabarti 1998, hereafter LMC98); bending instability of an accretion disc (Molteni et al. 2001a, hereafter M01a; Molteni et al. 2001b, hereafter M01b); interaction of accretion shocks with winds (Acharyya, Chakrabarti \& Molteni 2002, hereafter ACM02); and effect of cooling on the time dependent behavior of accretion flows (Chakrabarti, Acharyya \& Molteni 2004, hereafter CAM04). 

We base our studies of accretion onto neutron star mostly on MSC96, though we modify the SPH algorithm to suit our need to handle both hot and cold particles in the neutron star environment.

\subsection{Model Equations}
We consider a rotating, axisymmetric, inviscid flow around a neutron star. We consider the magnetic field to be negligible 
and ignore its effects completely. The gravitational force due to the compact object was modelled using the pseudo-Newtonian potential of Paczynski and Wiita (1980). The matter density ($\rho$), isotropic pressure ($P$) and the internal energy ($e$) of the flow are related to each other through $P=\rho e (\gamma-1)$. The adiabatic index $\gamma=4/3$, is kept constant throughout our simulations. We used a single-temperature model for the electrons and protons. The specific angular momentum $\lambda$ is varied from case to case but it is constant everywhere in a simulation setup, as the flow is purely inviscid. Furthermore, the SPH code uses
toroidal particles and thus it strictly preserves $\lambda$. The hydrodynamic code uses dimensionless quantities for computation. However, the cooling mechanisms require physical units (cgs is used here). For that purpose, all the relevant quantities are non-dimensionalized using their corresponding reference values. We use density of injected particles of the flow at the outer edge $\rho_{ref}=\rho_0$, the speed of light $v_{ref}=c$, and the Schwarzschild radius $x_{ref}=r_S=2GM_{NS}/c^2$ of the neutron star mass $M_{NS}$, as the reference density, velocity, and distance, respectively. From that we can derive the units of time $t_{ref}=x_{ref}/v_{ref}=r_S/c$, specific angular momentum $\lambda_{ref}=x_{ref}v_{ref}=cr_S$, mass $M_{ref}=\rho_{ref}x^3_{ref}=\rho_{ref}r_S^3$, and mass accretion rate $\dot{M}_{ref}=\rho_{ref}x^3_{ref}/t_{ref}=\rho_{ref}r_S^3c$.

We provide the Lagrangian formulae for the two-dimensional fluid dynamics equations for Smoothed Particle Hydrodynamics (SPH) in cylindrical coordinates, below.

The conservation of mass is given as (LMC98)

\begin{equation}
\frac{D\rho}{Dt}=-\rho \vec{\nabla} \cdot \vec{v}
\end{equation}

(here, $D/Dt$ is the comoving derivative).

The conservation of momentum is given by (LMC98, dropping the viscous terms from Eq. 2)
\begin{equation}
\frac{D\vec{v}}{Dt}=-\frac{1}{\rho} \vec{\nabla} P + \vec{g} + \frac{\lambda^2}{r^3}\hat{r}
\end{equation}

where, $\hat{r}$ is the radial direction vector and

\begin{equation}
\vec{g}=-\frac{1-\mathcal{C}}{2(R-1)^2}\frac{\vec{R}}{R},
\end{equation}

\begin{equation}
g_r=-\frac{1-\mathcal{C}}{2(R-1)^2}\frac{r}{R},
\end{equation}

\begin{equation}
g_z=-\frac{1-\mathcal{C}}{2(R-1)^2}\frac{z}{R}.
\end{equation}

Here, $\vec{R}=r\hat{r}+z\hat{z}$, $R=\sqrt{r^2+z^2}$, and $\mathcal{C}$ is the radiative pressure term arising out of the blackbody emission from the surface of the star. We have assumed the term to be isotropic for our simulations.

To achieve a better accuracy, we use a form of energy conservation where the sum of kinetic and thermal energies are used instead of only thermal energy (Monaghan 1985). Then, the energy conservation can be written as (following Eq. 9c of MSC96 and Eq. 11 of LMC98)

\begin{equation}
\frac{D}{Dt}\Big{(}e+\frac{1}{2}\vec{v}^2\Big{)}=-\frac{P}{\rho} \vec{\nabla} \cdot \vec{v} + \vec{v} \cdot \Big{(}\frac{D\vec{v}}{Dt}\Big{)} - \zeta_{1/2} \rho e^{\alpha}
\end{equation}

Here, $\zeta_{1/2}$ is the non-dimensional bremsstrahlung loss coefficient, as defined in MSC96,

\begin{equation}
\zeta_{1/2}=\frac{j \rho_{ref} x_{ref} T_{ref}^{1/2}}{c^3 m_p^2}
\end{equation}

and

\begin{equation}
T_{ref}=\frac{c^2 m_p \mu (\gamma - 1)}{k}
\end{equation}
where $\mu=0.5$ and $j=1.4 \times 10^{-27}$ cgs unit for ionized hydrogen (Allen 1973), $m_p$ is the mass of the proton, and $k$ is the Boltzmann constant. The subscript 1/2 to $\zeta$ signifies the use of a cooling law $\Lambda = \zeta_{1/2} \rho^2 e^{\alpha}$ with a constant $\zeta_{1/2}$ which is identical to the bremsstrahlung case ($\alpha=0.5$).

\subsection{SPH: Implementation of the cooling law}
We use the method described in detail by MSC96, while making changes to match the notations used so far. We have assumed the flow to be axisymmetric and thus, all the equations are written for cylindrical geometry. The interpolating kernel $W$ which is a function of cylindrical radial coordinate $\vec{R}$ and $k$th particle of mass $m_k$ as (MSC96)

\begin{equation}
m_k=2\pi \rho_k r_k \Delta \vec{R}_k.
\end{equation}

Any smooth function $A(\vec{R}_i)$ at position $\vec{R}_i$ can be defined as (MSC96),
\begin{center}
$A(\vec{R}_i)=\int A(\vec{R}) W(\vec{R}-\vec{R}_i;h) \frac{2\pi \rho r}{2\pi \rho r} d\vec{R}_k $
\end{center}

\begin{equation}
\approx \sum_k \frac{m_k}{2\pi \rho_k r_k} A(\vec{R}_k) W(\vec{R}_k-\vec{R}_i;h),
\end{equation}
where $h$ is the particle size. This simplifies the expression of the conservation laws and quantities that can be computed easily. As an example, the density at position $\vec{R}_i$ can be simplified as,
\begin{equation}
\rho(\vec{R}_i) \approx \sum_k  \frac{m_k}{r_k} W(\vec{R}_k-\vec{R}_i;h),
\end{equation}
which satisfies the continuity equation in cylindrical coordinates (MSC96).

The equations of motion to be solved using SPH reduces to three separate ones. The radial component of momentum equation,

\begin{equation}
\footnotesize{\Big{(} \frac{D v_r}{Dt} \Big{)}_i = \sum_k \frac{m_k}{r_k} \Big{(} \frac{P_i}{\rho^2_i} + \frac{P_k}{\rho^2_k} + \Pi_{ik} \Big{)} \frac{\partial W_{ik}}{\partial r_i} + \frac{\lambda^2}{r_i^3} - \frac{1-\mathcal{C}}{2(R_i-1)^2}\frac{r_i}{R_i},}
\end{equation}
the vertical component of the momentum equation,
\begin{equation}
\footnotesize{\Big{(} \frac{D v_z}{Dt} \Big{)}_i = \sum_k \frac{m_k}{r_k} \Big{(} \frac{P_i}{\rho^2_i} + \frac{P_k}{\rho^2_k} + \Pi_{ik} \Big{)} \frac{\partial W_{ik}}{\partial z_i} - \frac{1-\mathcal{C}}{2(R_i-1)^2}\frac{z_i}{R_i},}
\end{equation}
and the specific energy equation which can be written as,
\begin{equation}
\footnotesize{\Big{(} \frac{D (e + \vec{v}^2/2)}{Dt} \Big{)}_i = \sum_k \frac{m_k}{r_k} \Big{(} \frac{P_i}{\rho^2_i} + \frac{P_k}{\rho^2_k} + \Pi_{ik} \Big{)} - \frac{\Lambda_i}{\rho_i} + \vec{v}_i \cdot \Big{(} \frac{D\vec{v}}{Dt} \Big{)}_i .}
\end{equation}
where, $\Lambda_i = \zeta_{1/2} \rho_i^2 (e_i)^{\alpha} $ is the cooling term.

The kinematic dissipation is mimicked using artificial viscosities (MSC96), 
\begin{center}
$\Pi_{ij} = \frac{\alpha_v \mu_{ij} \bar{c}_{ij} + \beta \mu_{ij}^2 }{\bar{\rho}_{ij}}$, \\

$\mu_{ij} = \frac{r_i v_{ri} - r_j v_{rj}}{r_i (l_{ij}^2 + \eta_{ij}^2)} + \frac{(v_{zi} - v_{zj})(z_i -z_j)}{(l_{ij}^2 + \eta_{ij}^2)}  $, \\

$ l_{ij}^2 = (r_i -  r_j)^2 + (z_i - z_j)^2  $, $\eta_{ij} = \eta = 0.1 h^2$.
\end{center}
where, $\alpha_v$ and $\beta$ are the artificial viscosity coefficients.

The equations (12-14) have been adopted from MSC96 and LMC98, with the introduction of the term $\mathcal{C}$. It is to be noted that the kinematic viscosity modifies the energy significantly but does not modify the angular momentum beyond $0.05\%$ of the initial value. Thus, our simulations effectively reproduces the flow with constant angular momentum. The abbreviations for density $\bar{\rho}_{ij}$ and sound speed $\bar{a}_{ij}$ are taken from Monaghan (1992):
\begin{center}
$\bar{\rho}_{ij} = \frac{\rho_i + \rho_j}{2}$, $\bar{a}_{ij} = \frac{a_i + a_j}{2}$.
\end{center}

\subsection{Conservation of angular momentum}
We have used the following equation (LMC98 and MSC96) for the conservation of angular momentum,
\begin{equation}
\footnotesize{\Big{(} \frac{D v_{\phi}}{Dt} \Big{)}_i = - \Big{(} \frac{v_{\phi}v_r}{r} \Big{)}_i + \frac{1}{\rho_i}\Big{[} \frac{1}{r^2} \frac{\partial}{\partial r} (r^3 \mu_{ij} \frac{\partial}{\partial r} (\frac{v_{\phi}}{r}))  \Big{]}_i,}
\end{equation}
where, $\mu_{ij}$ is the kinematic viscosity. The terms $\alpha_v$ and $\beta$ control the amount of $\mu_{ij}$ necessary to reduce oscillations in shock transitions. However, the $2nd~term<<1st~term$ on the right hand side and made no significant contribution. We determined the value of $\lambda_i=r_i(v_{\phi})_i$ for the particle and use in Eqn. 12. It was observed that for all the cases we have tried, the determined angular momentum was almost constant and equal to the injected values up to an error of $0.05\%$. The average value over all particles deviated even less from the injected value, e.g. for C1 $\lambda_{avg}=1.800001$, matching the injected value up to 5 decimal places. This little numerical error is due to the fluctuation of the values of $(v_{\phi})_i$ and $(v_{r})_i$. Thus the obtained values of $\lambda_i$ are effectively equal to the $\lambda$ of the injected particles, which is obtained as a natural consequence of the conservation of angular momentum throughout the flow. As the flow is always sub-Keplerian, it can continue to the inner boundary which is the surface of the star. A part of the flow is absorbed into the star beyond $R=R_{NS}$ where the rotational velocity matches with that of the star and the surplus rotational kinetic energy is released through $\Delta \mathcal{E}(t)$ (see Boundary Conditions). As an example, the rate of transport of angular momentum onto the star for C1, i.e., the spin-up torque $N \approx \dot{M}_{accr} \lambda$ was such that the spin-up rate $\dot{\nu}=N/2\pi I$ was roughly around $5 \times 10^{-14}~Hz~s^{-1}$, where, the moment of inertia $I\approx 0.4 M_{NS} R_{NS}^2$. Thus any change of the spin of the star due to this feedback effect can be safely ignored for the purpose of calculations for the runtimes we have chosen.

\subsection{Boundary Conditions}

For our simulations, the pseudo-particles (or, just `particles' used in the text) are injected from $r_{inj}=30r_S$
with the same specific energy and specific angular momentum. The flow is assumed to be in vertical equilibrium when injected. The particles are tracked as long as they are within $r<r_{inj}$ and $R<R_{out}$, where, $R_{out}=35r_S$. The choice of $R_{out}$ was made so to minimize the computational time taken to track isolated particles moving far away from the inflow region. We have also carried out our simulations for a rectangular simulation box, but no significant changes were observed. For $C1$, injection velocity at equatorial plane with $v=0.1202$ and sound speed $a=0.0586$ was done which are appropriate for the transonic branch. Similarly, for cases $C2$ to $C5$, to ensure injection with same total energy and Mach number, we had to choose $v=0.1211$ and $a=0.0590$. In order to implement a reflection boundary condition at the inner boundary of $R_i < R_{NS} + h$ we used a reflecting condition for the velocity component $v_R$. For the $v_{\theta}$ component, a sliding or slipping boundary condition is used, where the flow maintains its $v_{\theta}$ value at the surface. The choice of boundary condition for $v_{\phi}$ component was based on both physical and numerical factors. We have tried multiple scenarios involving the presence and absence of absorption condition and the effect of no-slip and slip conditions with zero and non-zero viscosity. What we found was that, inviscid sub-Keplerian flows reaching the surface of the star would undergo the following processes: piling up at the boundary $\longrightarrow$ forming a highly dense layer $\longrightarrow$ a local increase in viscosity $\longrightarrow$ redistribution of the angular momentum to match the $v_{\phi}$ of the layer with that of the star $\longrightarrow$ being absorbed, effectively. Numerically, the most efficient scheme for inviscid flows turned out to be the one with no-slip and absorption, after reaching the surface, which implicitly takes care of the mentioned steps. The explicit study of viscous flows, with different inner boundary conditions and the transition mentioned above, are beyond the scope of the present work and are to be reported in a separate paper as stated earlier. Thus, for the cases reported here, a no-slip condition is used for the azimuthal component where the $v_{\phi}$ of the flow is matched with the angular velocity of the star ($\omega_{NS}r_{NS}$) at the surface. For our calculations, $\omega_{NS}=142 Hz$. Along with the cooling criteria, these conditions allow matter to settle down on the surface of the star and also allows meridional motion from the equatorial region towards the poles (and vice-versa). However, if the flow reaches the $R_i = R_{NS}$, it is immediately absorbed and all the thermal and kinetic energy of the particle is assumed to be released as blackbody radiation. In the present simulations we did not study the emission due to Comptonization of seed photons originating from bremsstrahlung or blackbody radiation. We, however, included the effects of radiation pressure on the flow through the quantity $\mathcal{C}$. We report the cases where the temperature is self-consistently modified by taking into account the additional flux arising out of energy of the accreted particles.
The initial surface temperature $T^0_{NS}$ of the neutron star was kept constant for all cases ($C1$ to $C5$) at $T^0_{NS}=0.0~keV$. The term $\mathcal{C}$ is controlled by the total energy deposited by the particles at the surface of the star ($\Delta \mathcal{E}$) and it given by,
\begin{equation}
\mathcal{C}(t)=\frac{T_{NS}^4(t)\sigma_{bb} R_{NS}^2 m_p}{c \sigma_T},
\end{equation}
where $T_{NS}(t)=\Big{(}(T_{NS}^0)^4+\frac{\Delta\mathcal{E}(t)}{\sigma_{bb} 4 \pi R_{NS}^2 dt}\Big{)}^{0.25}$. Here, $\sigma_{bb}$ is the Stefan-Boltzmann constant, $m_p$ is the mass of proton, $\sigma_T$ is the Thompson scattering cross section. 

\subsection{Coalescence of Particles}

One of the major problems with the previous version of the code was that it did not dynamically evolve when the particles came very close to each other ($R_{ij} < h$) and the resulting hydrodynamical timescale became very small. This problem is even more accute in our case as particles tend to aggregate on the surface of the star. In the original code, within $150$ time steps the value of $dt < 0.0002$ and effectively stopped the evolution. A simple absorption condition at the boundary did not solve the problem as the aggregates grew beyond the surface's immediate vicinity. To circumvent this, we implemented a 2-particle coalescing scheme (adopted from Vacondio et al. 2013) when the inter-particle distance $R_{ij} < 0.1 h$.

After each time step, we check for neighbours using a $linked-list$ algorithm. The neighbour list of a particle $i$ is then searched for all such neighbours $j$, for which $R_{ij} < 0.1 h$. The minimum of such $R_{ij}$ values and the corresponding $(i,j)$ pair is selected. If such pairs are found, a list of such pairs is made and the following scheme is applied to coalesce particle-pairs $(i,j)$.

To conserve the total mass:
\begin{equation}
m_k=m_i+m_j, k=min(i,j).
\end{equation}
We preserve the total momentum, by using:
\begin{equation}
\vec{v}_k=\frac{m_i\vec{v}_i+m_j\vec{v}_j}{m_i+m_j}.
\end{equation}
We also preserve the total thermal energy, by using:
\begin{equation}
e_k=\frac{m_i e_i + m_j e_j}{m_i+m_j}.
\end{equation}

The new location of the particle is determined by,
\begin{equation}
\vec{R}_k=\frac{m_i\vec{R}_i+m_j\vec{R}_j}{m_i+m_j}.
\end{equation}
After going through the list of all such pairs, a standard re-indexing is done for all the particles still left in the system. As a result, the average number of particles present in the simulation stayed between $\sim 14000 $ and $\sim 16000$ for $C1$ and between $\sim 11500$ and $\sim 13000$ for $C2-C5$.

\begin{table*}
\footnotesize{
\vskip0.2cm
\begin{center}
\caption{Parameters for the simulations and centroid frequencies (in Hz) found in the PDS of bremsstrahlung loss.}
\vskip0.2cm
 \begin{tabular}{c c c c c c c c c c}
 \hline
 \hline
 ID & $\dot{m}_h$ ($\dot{M}_{EDD}$) & $\lambda$ ($r_Sc$) & $R_{NS}$ ($r_S$) & $\alpha$ & $\nu_{LF1},Q_{LF1}$ & $\nu_{LF2},Q_{LF2}$ & $\nu_{hHz},Q_{hHz}$ & $\nu_{l},Q_{l}$ & $\nu_{u},Q_{u}$ \\  
 \hline
 \hline
 C1 & 0.094 & 1.8 & 3.0 & 0.5 & - & - & 128.88, 17.82 & 213.31, 3.17 & - \\ 
 C2 & 0.094 & 1.7 & 3.0 & 0.5 & - & - & 102.22, 3.91 & 425.40, 5.61 & 633.78, 1.77 \\ 
 C3 & 0.188 & 1.7 & 3.0 & 0.5 & - & - & 115.55, 5.02 & 486.08, 2.49 & 742.73, 1.48 \\ 
 C4 & 0.094 & 1.7 & 3.0 & 0.6 & 41.62, 1.30 & 94.41, 4.66 & 137.58, 5.78 & 434.97, 4.96 & 599.23, 3.51 \\ 
 C5 & 0.094 & 1.7 & 4.0 & 0.5 & 36.36, 2.91 & 77.55, 3.14 & 196.11, 18.06 & 465.70, 5.11 & 654.67, 3.51 \\ 
 \hline
\end{tabular}
\label{table:parameters}
\end{center}
}
\end {table*}

\subsection{Timing Analysis}
We computed the total emitted energy due to bremsstrahlung $\mathcal{E}(t)$ by integrating the emission per unit mass ($\Lambda_i/\rho_i$) over the particles existing in the simulation $(n)$,

\begin{equation}
\mathcal{E}(t)=\sum_{i=1}^{i=n} \Lambda_i m_i/\rho_i.
\end{equation}

As a consequence, the densest and hottest regions contributed more to the time variation of the bremsstrahlung loss. In order to probe the hydrodynamic characteristics of CENBOL and NBOL, the emitted energy was plotted against time to generate the lightcurve. To extract periodic features, we used NASA's \textsc{ftools} package to create the fast Fourier transformed Power Density Spectra (PDS) from the lightcurves. Data was gathered after every $10^{-5}~s$ to generate the lightcurve. We used \textsc{powspec} command, with a rebinning factor of -1.05 for all the cases to generate the PDS. Any oscillatory signature (such as a Quasi Periodic Oscillation or a peaked noise) is reflected as a peak in the PDS. We report only those 5 cases (see, Table 1.) where significant variations of QPOs are observed. To avoid the effects of the transient phase, we only used data collected from $0.05~s$ to $0.3286~s$. A Lorentzian profile is used to fit a QPO in the PDS. We determined the centroid frequencies, full-width half maxima and rms power up to $90\%$ confidence level. While fitting the QPOs with Lorentzians, we only fitted the fundamental frequency of a type of QPO, when the harmonic is overlapping with another fundamental frequency. In case of such an overlap we take the best statistical fit (one peak or a mean position as the case demanded). When no such overlap occurs, and the harmonic feature is significant, the corresponding Lorentzian is fitted and mentioned in the text (Table 1).  

\section{Results}
We studied multiple cases (see, Table 1) by varying specific angular momentum ($\lambda$), radius of the star ($R_{NS}$), accretion rate ($\dot{m}$, alternatively used as the halo accretion rate $\dot{m}_h$), cooling index ($\alpha$), one at a time, to study the effects in accretion and outflow. 

We first discuss a typical scenario in detail to highlight the hydrodynamical aspects of the flow.

\textbf{Case C1:}
In order to demonstrate hydrodynamic properties, we first focus on case $C1$. The initial simulation parameters are mentioned in the Table 1. The simulation was run for $32860$ time steps ($t=0.3286 s$), which is equivalent to $200$ dynamical time steps at $30~r_S$. The cooling process was switched on after ($t \sim 0.0083 s$). The system settled into a steadily oscillating state after a brief transient phase (up to $t \sim 0.0500 s$). Indeed, the temperature of NBOL was found to fluctuate around $0.15$keV even though
initially its value was chosen to be $\sim 0.0$keV. Figure 1(a) shows variation of NBOL temperature during our simulation. This
would induce fluctuations of the NBOL height as well. Details of the behaviour of NBOL is beyond the scope of the current paper and would be published elsewhere. 
In this case, we studied PDS of the mass outflown through the upper quadrant. The hHz frequency corresponding to vertical oscillation was found at $\nu_{hHz}=127.22~Hz$, with a harmonic at $\nu_{hHz}=254.40~Hz$. For black holes, Molteni et al. 2001 found such oscillations due to bending instabilities in the flow. When scaled for the mass of the neutron star (here, assumed to be $M_{NS}=1.0M_{\odot}$ for computational purposes), the frequency range comes out as 20-300 Hz. Thus, we can clearly identify the oscillations in the lightcurve as well as the mass outflow due to such instabilities, which is also reflected in Fig. 2(c-d).\\

\begin{figure*}[ht]
  \centering
\includegraphics[height=6.0cm,width=6.0cm]{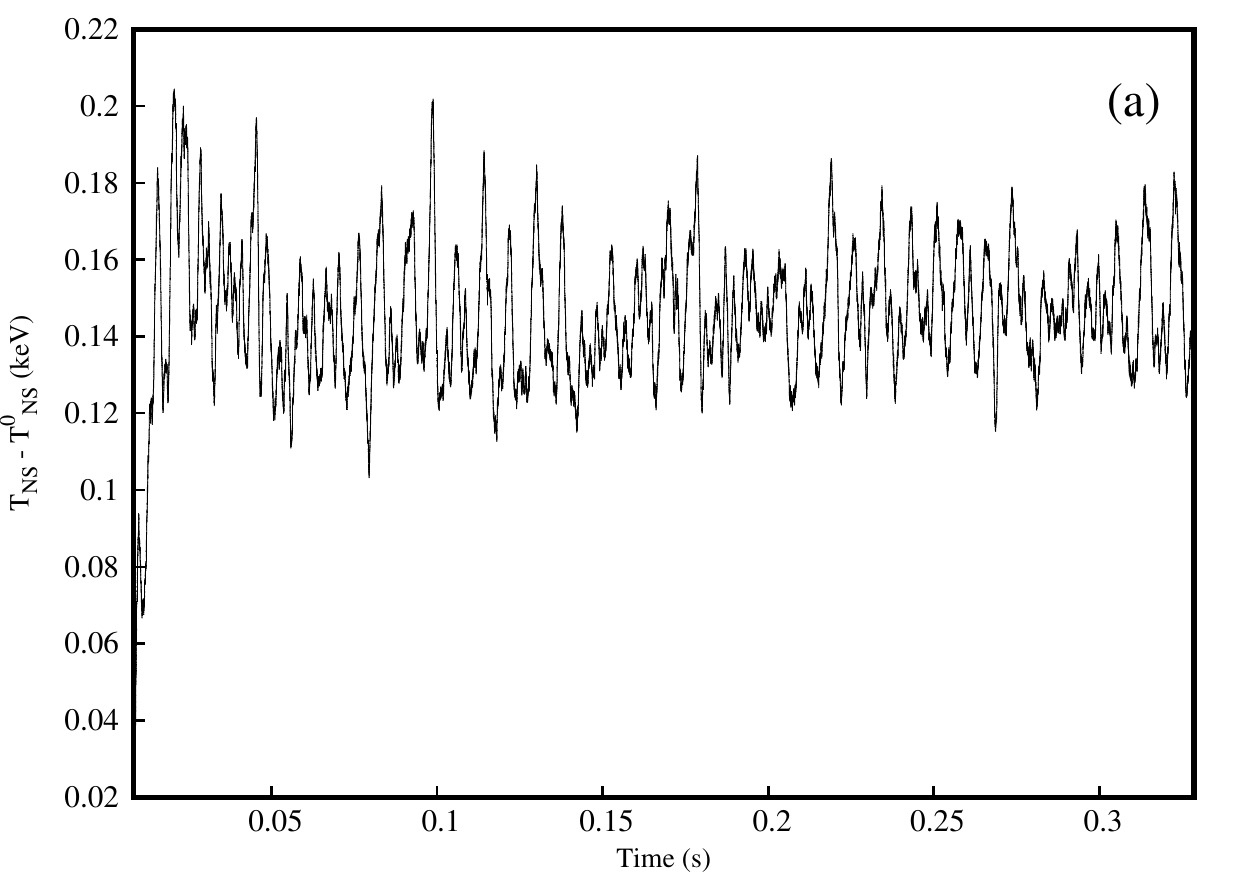}
\includegraphics[height=6.0cm,width=4.0cm]{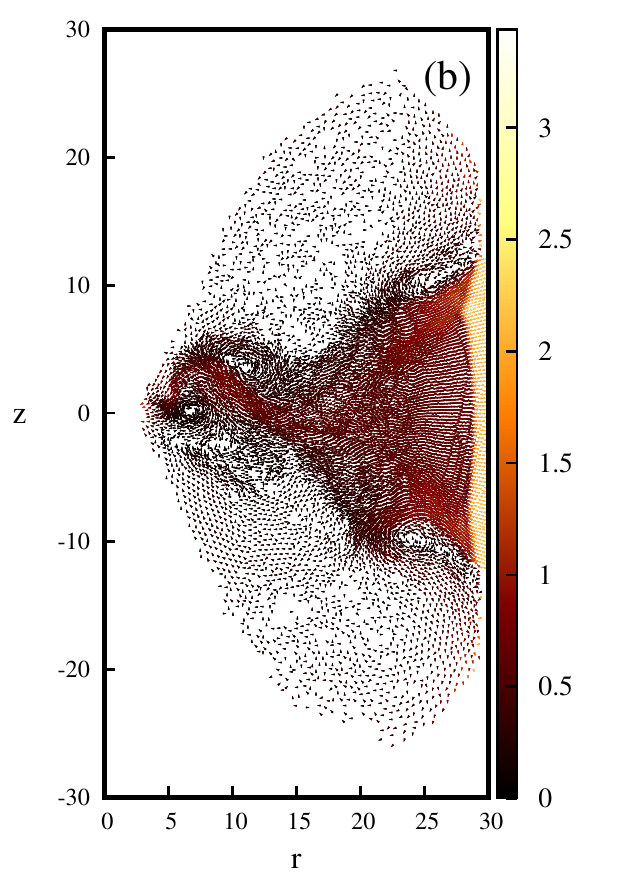}
\includegraphics[height=6.0cm,width=6.0cm]{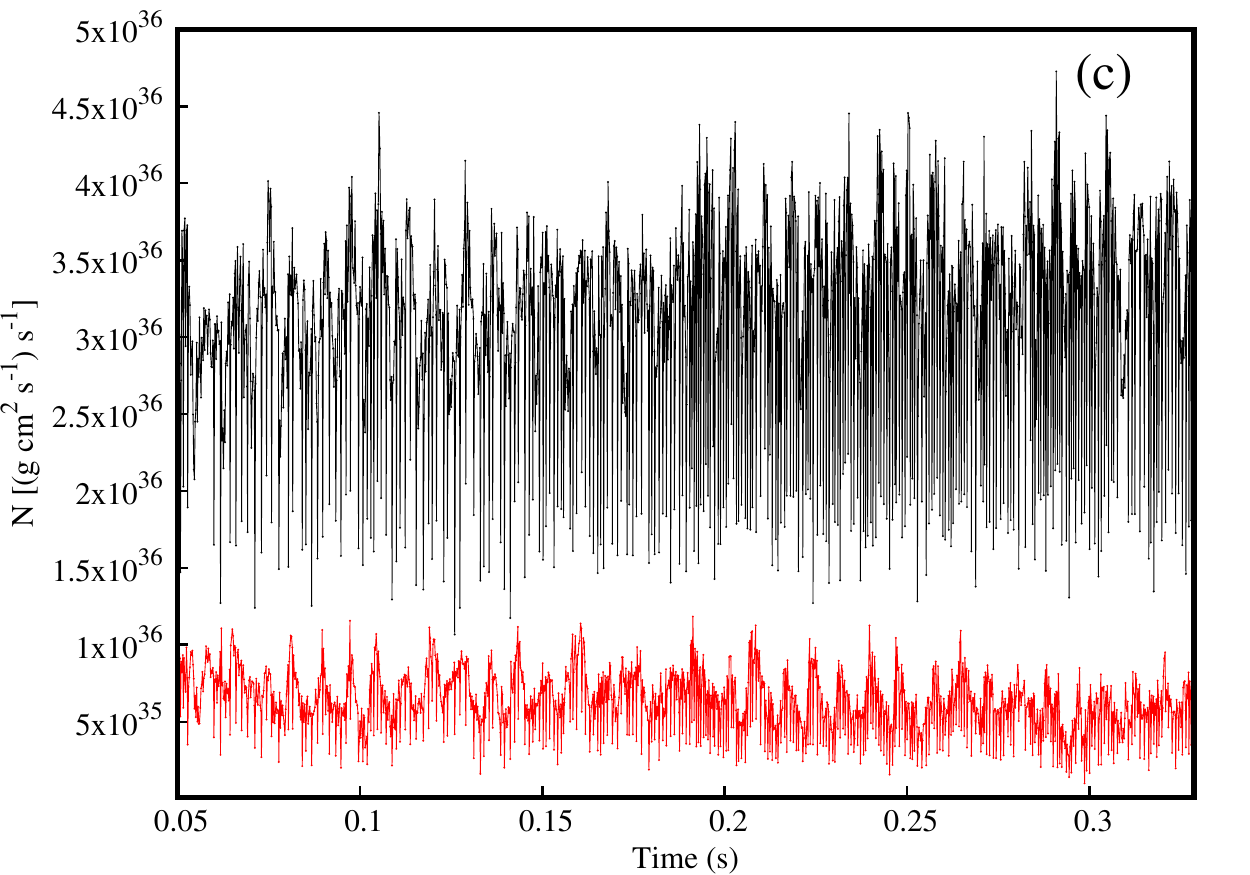}
\caption{(a) The variation of $T_{NS}$ with time (s), (b) the velocity vector $v_r \hat{r} + v_z \hat{z}$ (arrow heads) with the Mach number in the colour bar, for the flow configuration $C1$ at time $t=0.2690 s$, (c) variation of the rate of transfer of angular momenta (N) onto the surface of the star (black) and in the outflows (red) with time (s).}
\label{fig1}
\end{figure*}

	In Fig. 1(b), the velocity in $r-z$ plane is plotted for all the particles. The Mach Number values are shown in the colour gradient. Multiple turbulent cells are seen to be formed due to the interaction of wind and accreting matter. The formation of outflow, in both upper and lower quadrants, from the post-shock region of CENBOL is also captured. In the lower quadrant, a portion of the outflow falls back on to the inflow as a feedback.

In Fig. 1(c), the variation of the rate of transfer of angular momentum (N, see sect. 3.3) with time is plotted for accretion (black) and outflows (red). Given that the specific angular momentum of the flow remains constant, the variation of the transfer rate with time depends on the mass flow rate in accretion and outflow, respectively. As the flow is always sub-Keplerian, the solution allows the matter to fall onto the star and transfer the angular momentum on the surface of the star. However, the spin-up torque $(N)$ plotted in the Figure was such that it corresponded to a spin-up rate of $\sim 5.0 \times 10^{-14} Hz~s^{-1}$. The value agrees well with observational results and predictions from other models (Bildsten 1998; Revnivtsev \& Mereghetti, 2015; Sanna et al. 2017; Bhattacharyya \& Chakrabarty, 2017; G{\"u}gercino{\v g}lu \& Alpar, 2017; Ertan 2018). The rest of the angular momentum is carried out by the outflows in both quadrants.

In Fig. 2(a), we show the density contours [$log(\rho/\rho_0)$]. The temperature contours (logarithmic scale) are plotted in Fig. 2(b). The contours of constant Mach number for accreting matter, at time $t=0.2690 s$, is shown in Fig. 2(c). The panel 2(d) shows the Mach number contours for the same case C1, at time $t=0.2528 s$. 
The flow puffs up after the centrifugal pressure supported shock (CENBOL) and acts as the base of the outflowing matter. Matter nearest to the star is observed to be hottest and densest. A dense and hot outflow is seen to emerge from within the CENBOL region. The Mach number contours show that the flow has time-dependent asymmetric distribution about $z=0$ plane. For C1, the high angular momentum creates the outer shock at around $26~r_S$ on the equatorial plane. The subsonic flow beyond the shock surface slowly becomes supersonic near the inner boundary of the star and finally settles on the surface through a strong shock. Notice that the Mach number contours near the edge of the star show supersonic behaviour of the flow
and no contours of subsonic Mach number are plotted. This is due to the fact that in the plotted case, 
the shock surface was very close to the boundary and the subsonic pseudo-particles were absorbed at the surface, 
before being written out, within the code. 

\begin{figure*}
\includegraphics[height=7.0cm,width=4.0cm]{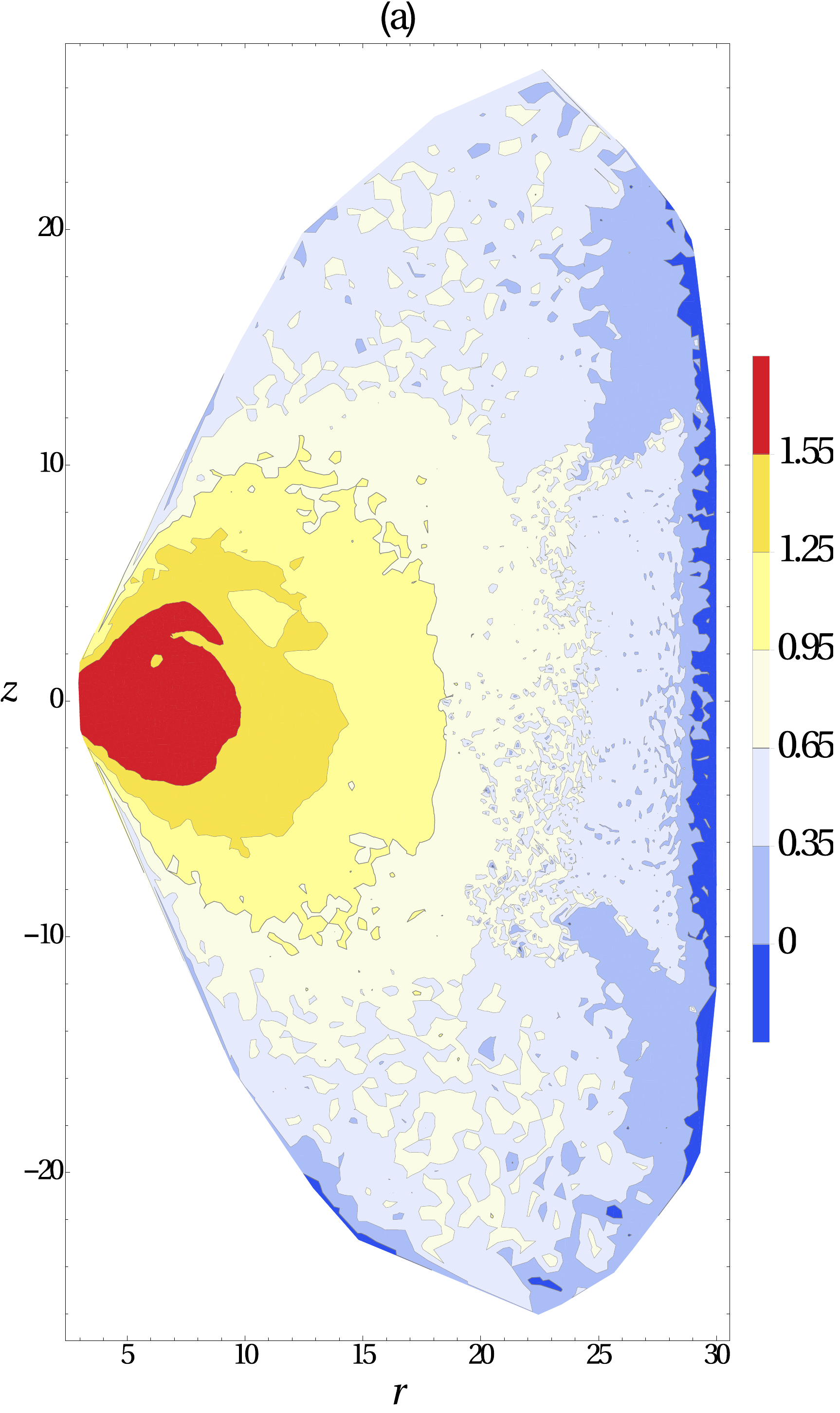}
\includegraphics[height=7.0cm,width=4.0cm]{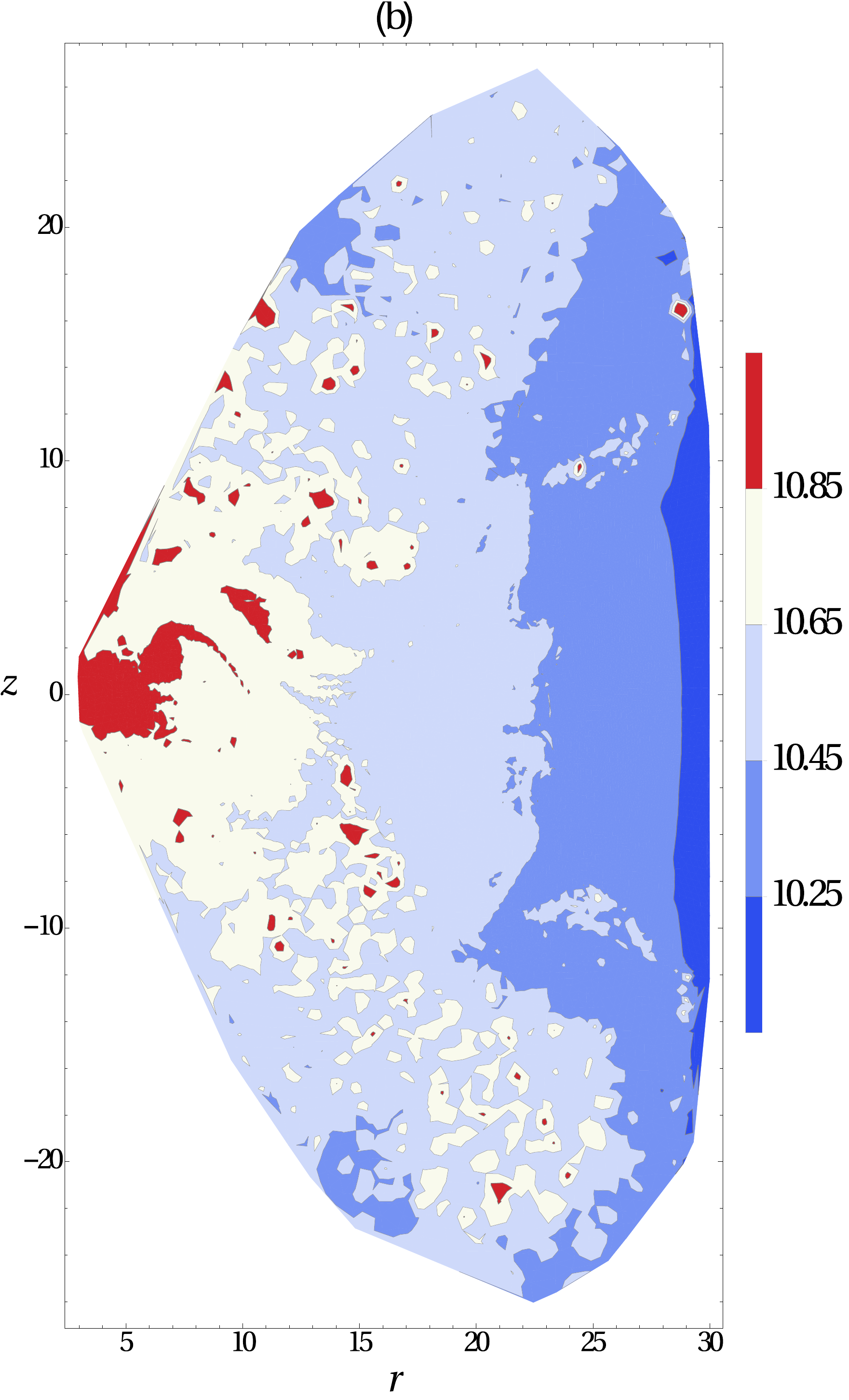}
\includegraphics[height=7.0cm,width=4.0cm]{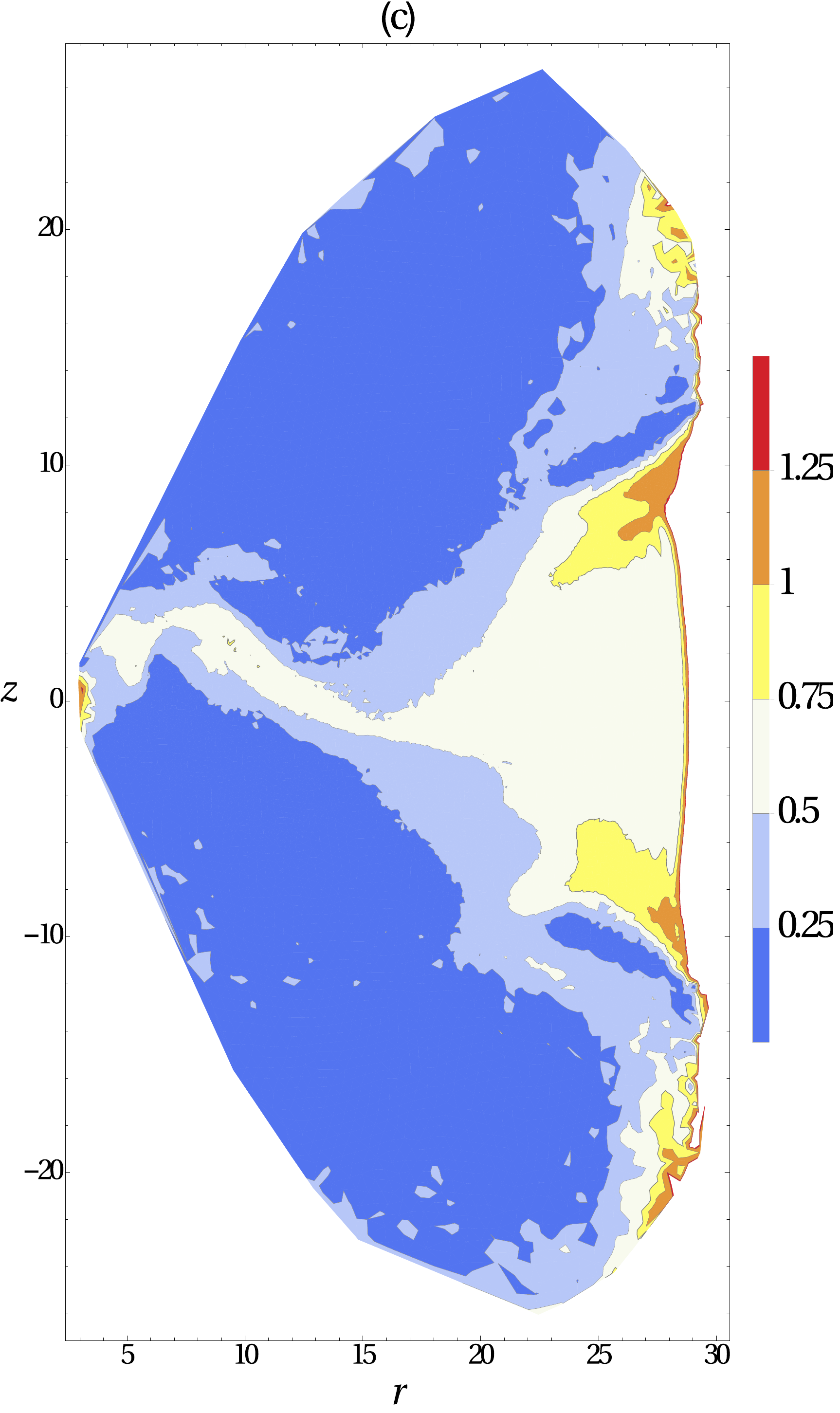}
\includegraphics[height=7.0cm,width=4.0cm]{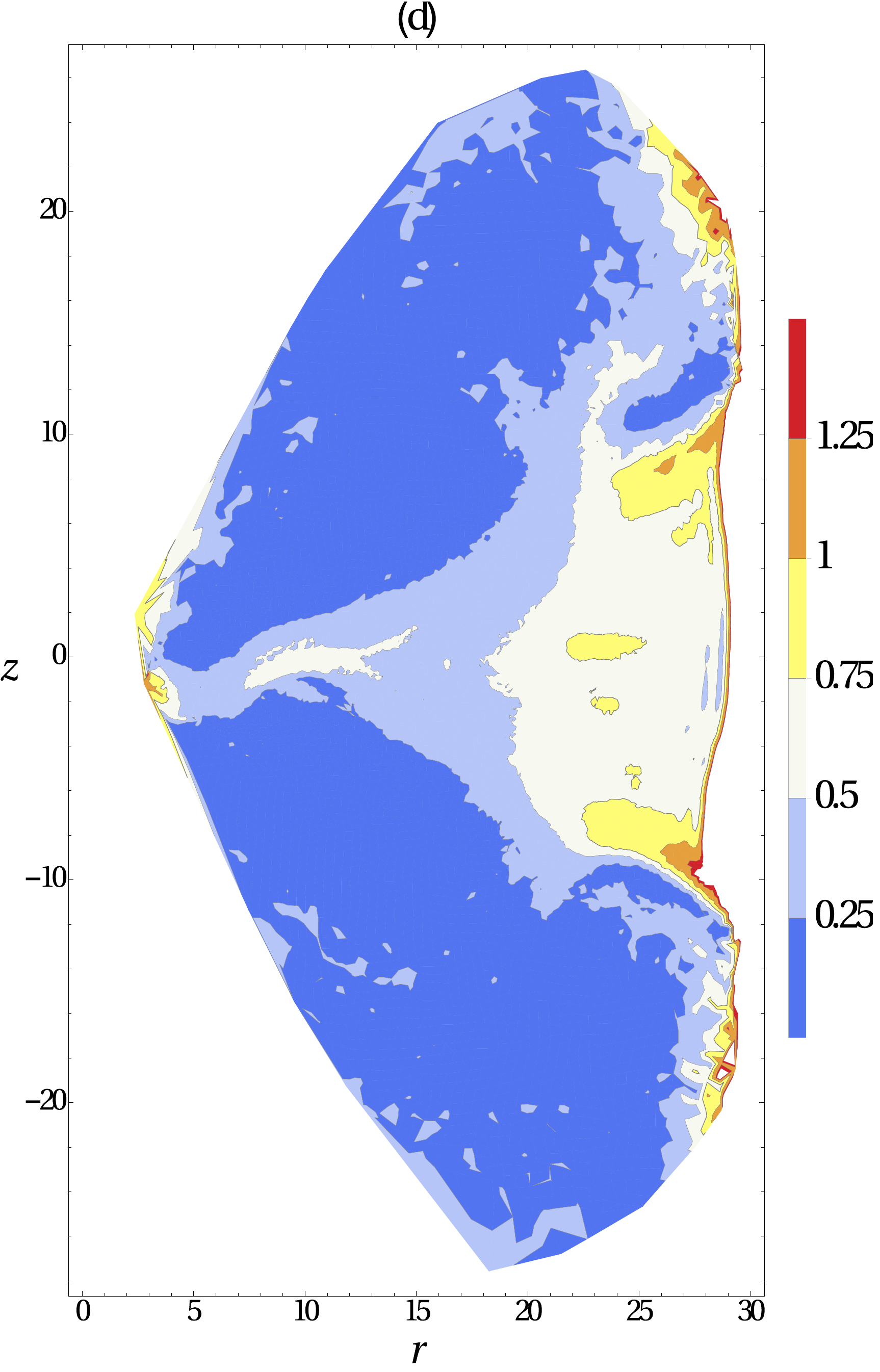}
\caption{The flow configuration for the case $C1$: (a) $log(\rho/\rho_0)$, (b) log of temperature in $K$, (c) Mach number contours for accreting inviscid flow, at $t=0.269 s$  and at(d)  $t=0.2528 s$. Both the density and temperature change by an order of magnitude in the inner part of CENBOL, as compared to the outer part. The flow configuration in (c) shows only the pre-shock region of the secondary shock (NBOL) at the boundary. However, in (d), we see that the shock has formed at $\sim 5~r_S$, and a shock in outflow forms at very close to the surface of the star. The outflow (in both (c) and (d)) becomes supersonic near the outer edge of the simulation boundary.}
\label{fig1}
\end{figure*}

\textbf{Case C2:}
When $\lambda$ is reduced from $1.8$ to $1.7$, it reduces the asymmetry around $z=0$ plane (Fig. 3a). The shock near the star becomes more prominent and the outflow profile changes. For $C1$, most of the outflow is generated from the immediate vicinity of the post-shock region of the CENBOL. We plot the ratio of the mass outflow to the mass accreted onto the star in unit time (denoted by $\dot{M}_{out}/\dot{M}_{accr}=M_{out}/M_{accr}$) for both $C1$ and $C2$ in Fig. 3(c). Apart from the initial higher values for $C1$, the ratio is comparable. However, the particle coalescing scheme merges the particles in higher density region, which reduces the number of particles in the simulation. The inset panel of Fig. 3(c) shows the same comparison in terms of the ratio of number of particles outflown to the number of accreted particles in unit time (denoted by $\dot{n}_{out}/\dot{n}_{accr}=n_{out}/n_{accr}$). The lower value for $C2$ further affirms the fact that bulk of the outflow is generated near the NBOL for $C2$. The outflow in Fig. 3(a) also undergoes a shock transition before becoming transonic again. In Fig. 4 (c), we plot the PDS of bremsstrahlung loss for the case C3. As $\lambda$ was decreased from $1.8$ to $1.7$, the centrifugal pressure dominated force is decreased. However, the flow was injected with same total energy as that of case C1 which resulted in a higher radial velocity. This increases the ram pressure of the fluid flow and prompts the resonance oscillation to take place at a smaller radius. The resulting oscillations are also seen to be dominated by the radial motion as compared to the vertical motion. The CENBOL moves closer to the star surface. The hecto-Hz QPO is still present and observed at $\nu_{hHz}=102.22~Hz$. We also observe the twin kHz QPOs between $400~Hz$ and $700~Hz$. The centroid frequencies of the lower and upper kHz QPOs are at $\nu_{l}=425.40~Hz$ and $\nu_{u}=633.78~Hz$, respectively. 

\begin{figure*}
  \centering
\includegraphics[height=6.0cm,width=3.5cm]{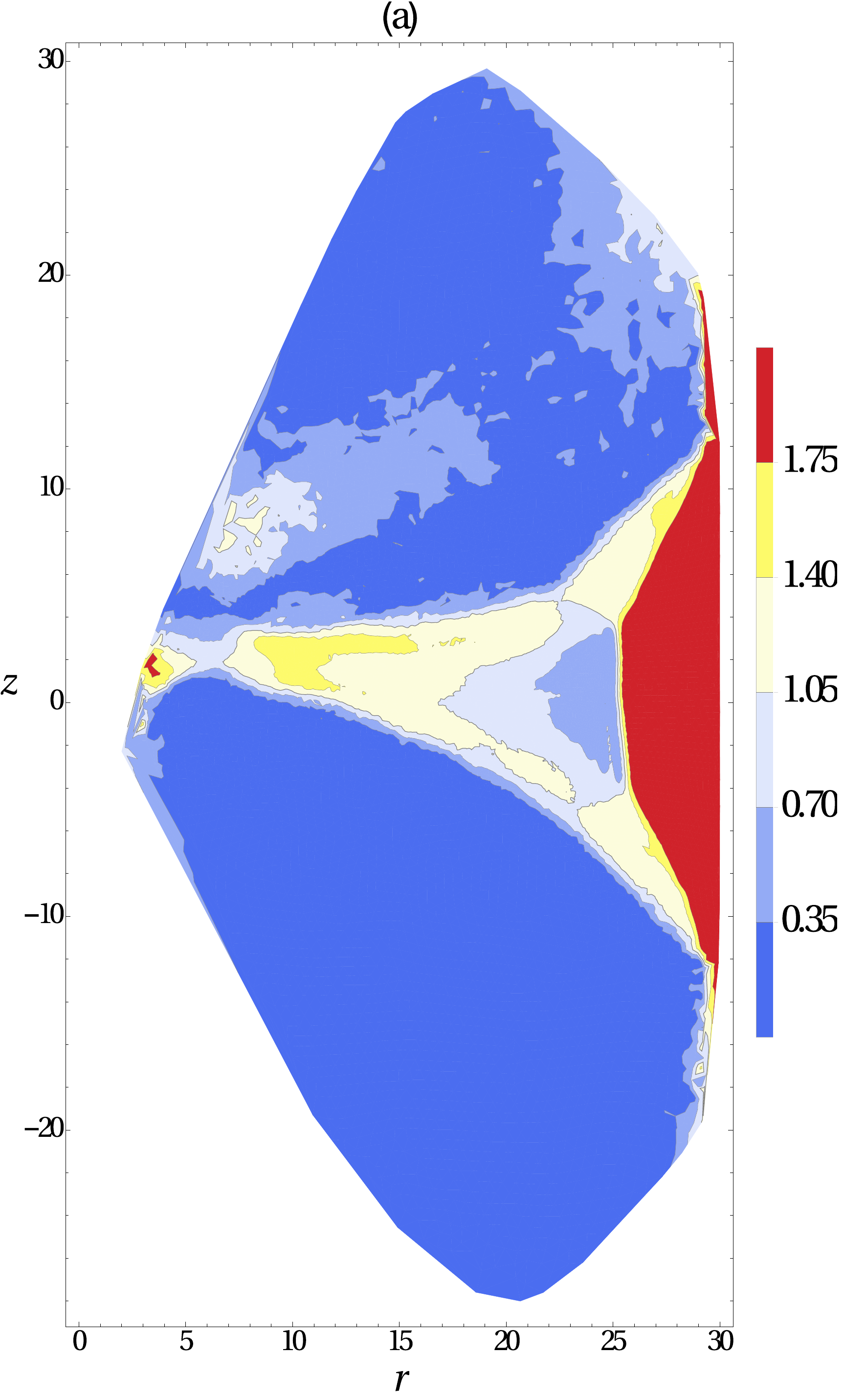}
\includegraphics[height=6.0cm,width=6.0cm]{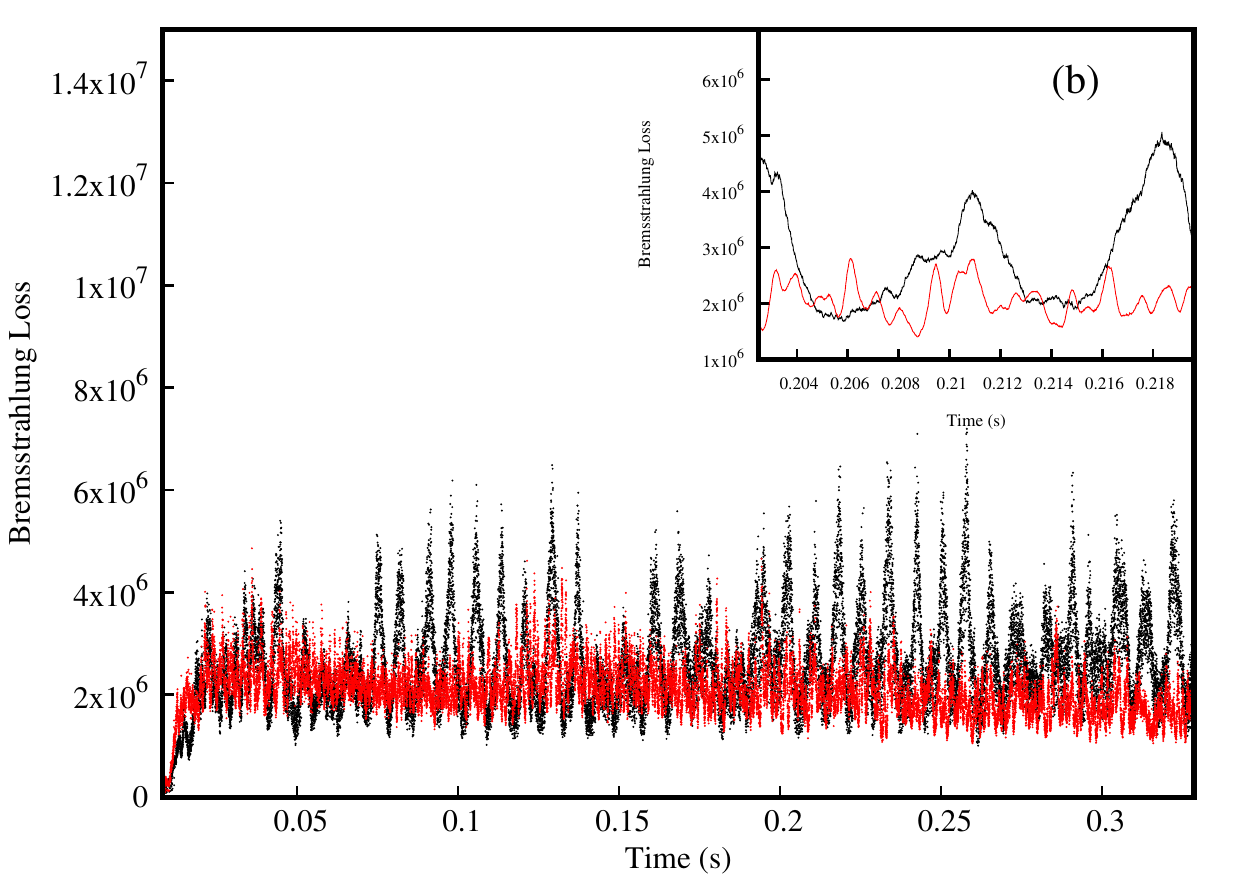}
\includegraphics[height=6.0cm,width=6.0cm]{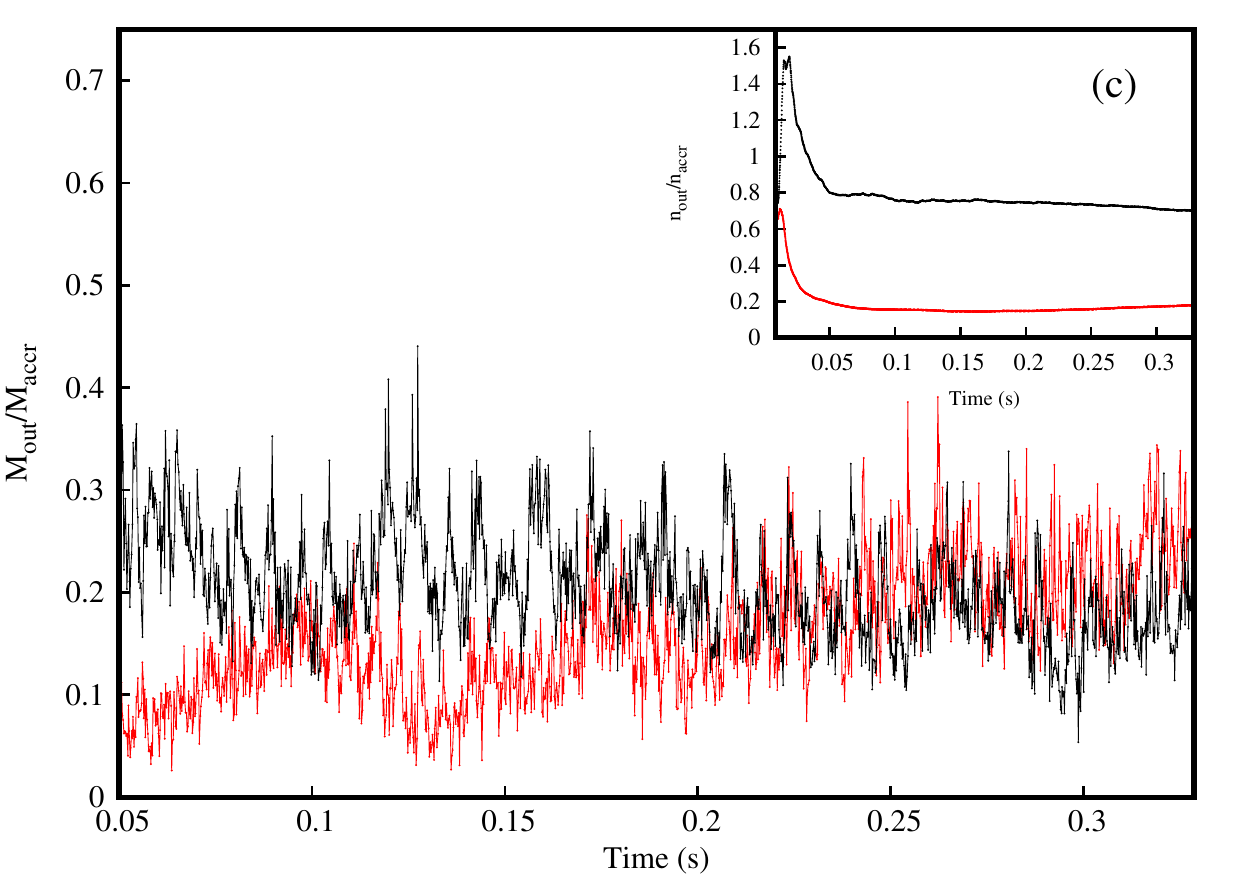}
\caption{(a) Contours of constant Mach Number for the case $C2$, at time $t=0.269 s$. The outer shock forms closer to the star at $(\sim 25~r_S)$. The inner shock is formed at $\sim 5~r_S$. The shock in outflow is diffused as compared to $C1$ and is formed between $\sim 10-15~r_S$. The outflow became transonic near the outer edge. In (b), we show the lightcurves of the total bremsstrahlung loss from the system for cases $C1$ (black) and $C2$ (red). Both the lightcurves show periodic behavior with distinctly different frequencies. In (c), the ratio $M_{out}/M_{accr}$ ($=\dot{M}_{out}/\dot{M}_{accr}$) for both $C1$(black) and $C2$(red) is plotted. The inset of (c) shows the $n_{out}/n_{accr}$ ratio ($=\dot{n}_{out}/\dot{n}_{accr}$), which shows more particles were coalesced near the NS boundary and the outflow generated from the regions closer to the NS surface had a low $n_{out}/n_{accr}$ value despite having comparable $M_{out}/M_{accr}$ values.}
\label{fig1}
\end{figure*}

\textbf{Case C3:}
In Fig. 4 (d), we plot the PDS of bremsstrahlung loss for C3. As $\dot{m}_h$ was increased from $0.094$ to $0.188$, the effective radiative cooling due to the bremsstrahlung process is also increased (higher $\rho$). The decrease in cooling timescale, prompts the resonance oscillation to take place at a smaller radius. This brings the NBOL closer to the surface. The hecto-Hz QPO is also observed at $\nu_{hHz}=115.55~Hz$. Both the lower and upper kHz QPOs show an increase in centroid frequency (compared to C2) at $\nu_{l}=486.08~Hz$ and $\nu_{u}=742.73~Hz$, respectively. 

\textbf{Case C4:}
As $\alpha$ was increased from $0.5$ to $0.6$, the effective radiative cooling due to the bremsstrahlung process decreased (lower $\Lambda$). The increase in cooling timescale, prompts the resonance oscillation to take place at a larger radius. This pushes the NBOL away from the surface. The PDS of bremsstrahlung loss for the case C4 is shown in Fig. 4(e). The low-frequency QPOs are observed at $\nu_{LF1}=41.62~Hz$ and at $\nu_{LF2}=94.41~Hz$. The hecto-Hz QPO is also observed at $\nu_{hHz}=137.58~Hz$. Both the lower and upper kHz QPOs are also observed at $\nu_{l}=434.97~Hz$ and $\nu_{u}=599.23~Hz$, respectively. 

\textbf{Case C5:}
In this case, as $R_{NS}$ was increased from $3~r_S$ to $4~r_S$, the effective radiative pressure due to the blackbody emission from the surface of the star decreased (lower $\mathcal{C}$). This aided the gravitational force in bringing both the shocks closer to the surface. The reduction of the outer edge of CENBOL increased the centroid frequency, above $10~Hz$, corresponding to the low-frequency QPO $\nu_{LF1}=36.36~Hz$ and its harmonic $\nu_{LF2}=77.55~Hz$, making them detectable (Fig. 4(f)). The hecto-Hz QPO is also observed at $\nu_{hHz}=196.11~Hz$. Both the lower and upper kHz QPOs are also observed at $\nu_{l}=465.70~Hz$ and $\nu_{u}=654.67~Hz$, respectively. 

\begin{figure*}[ht]
  \centering
\includegraphics[height=4.0cm,width=4.5cm,angle= 0]{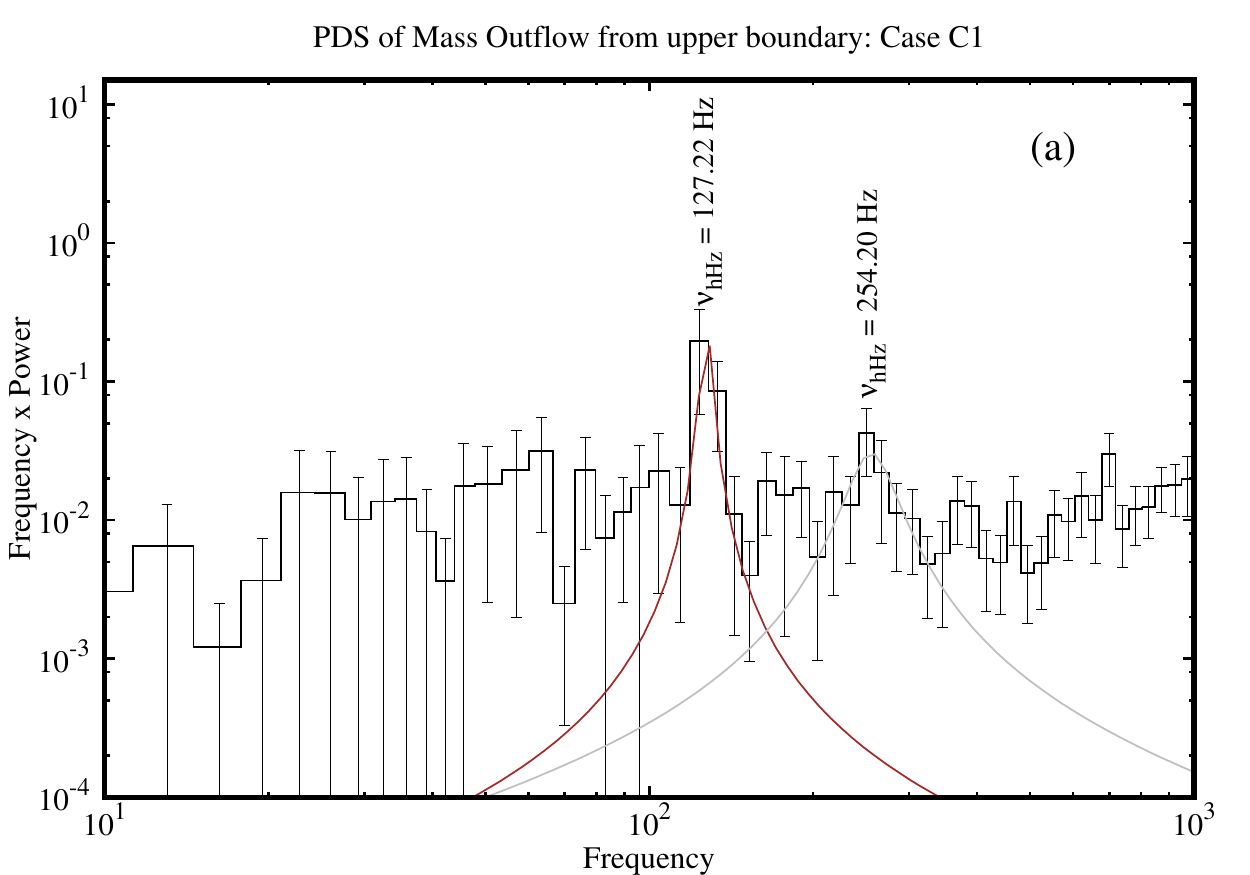}
\includegraphics[height=4.0cm,width=4.5cm]{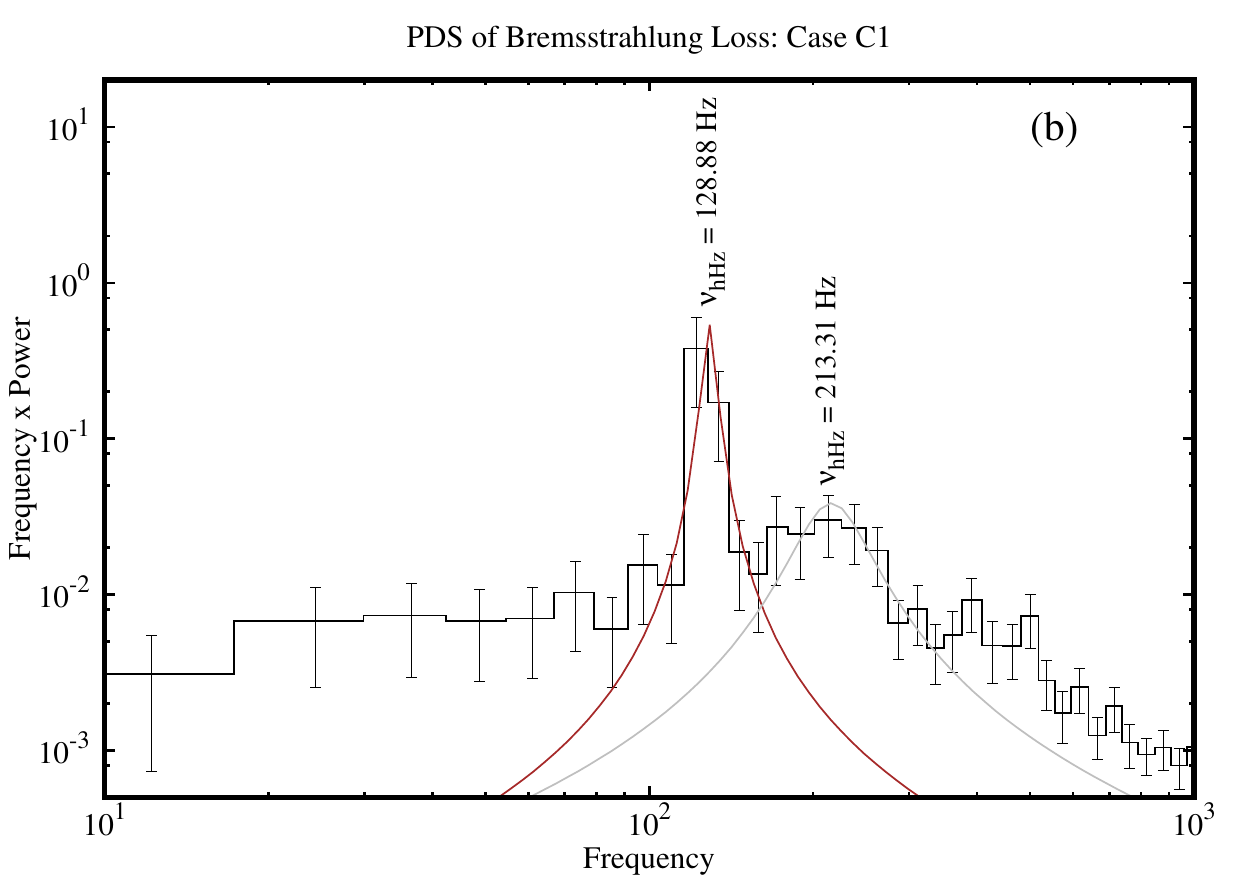}
\includegraphics[height=4.0cm,width=4.5cm]{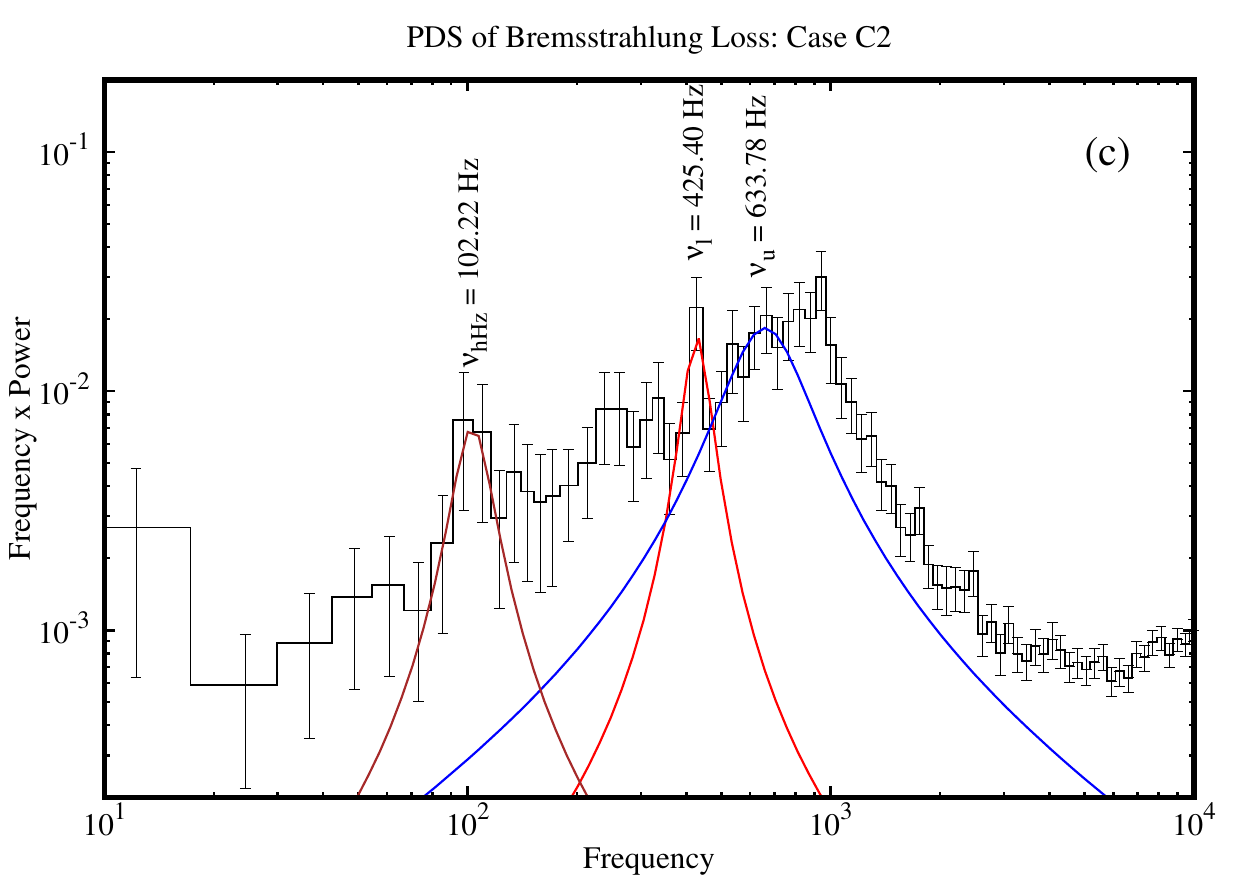}
\includegraphics[height=4.0cm,width=4.5cm]{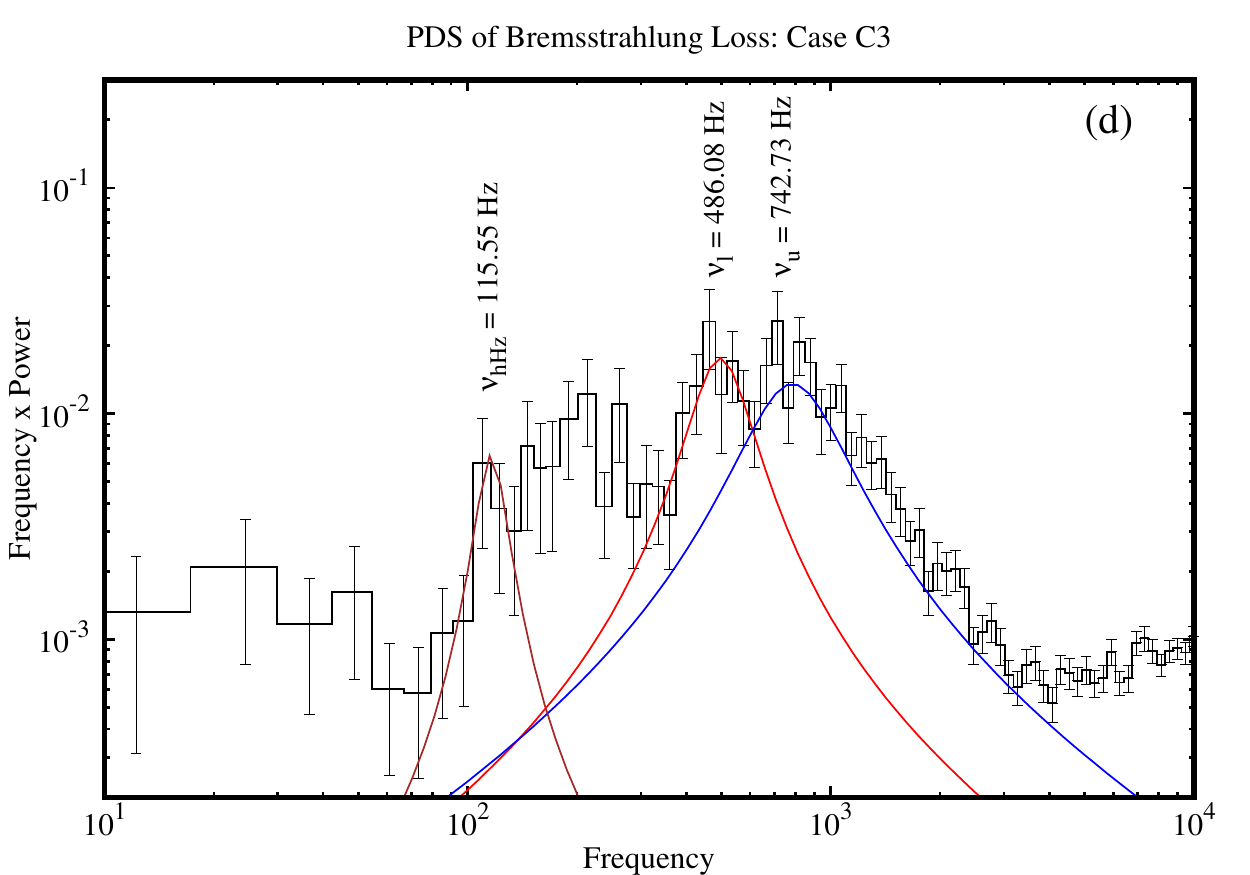}
\includegraphics[height=4.0cm,width=4.5cm]{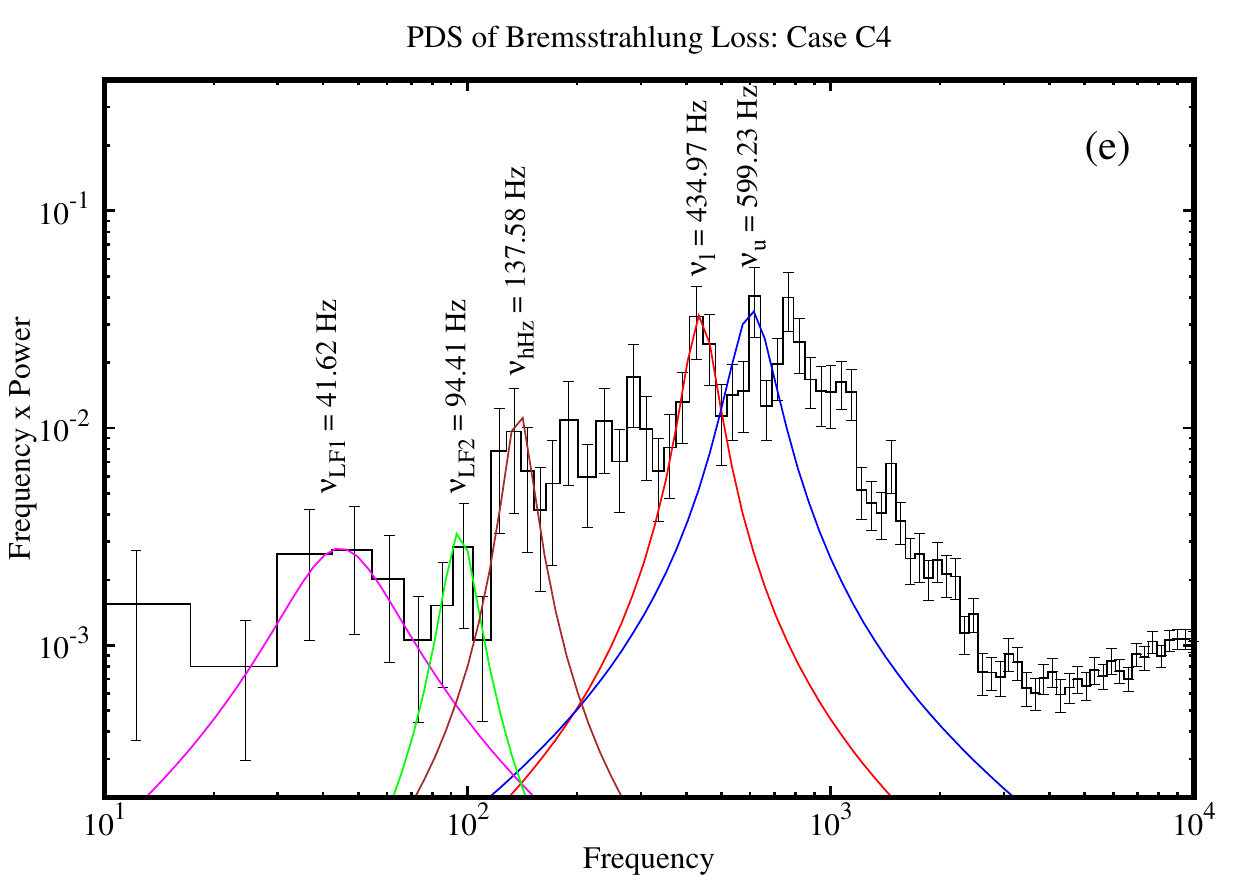}
\includegraphics[height=4.0cm,width=4.5cm]{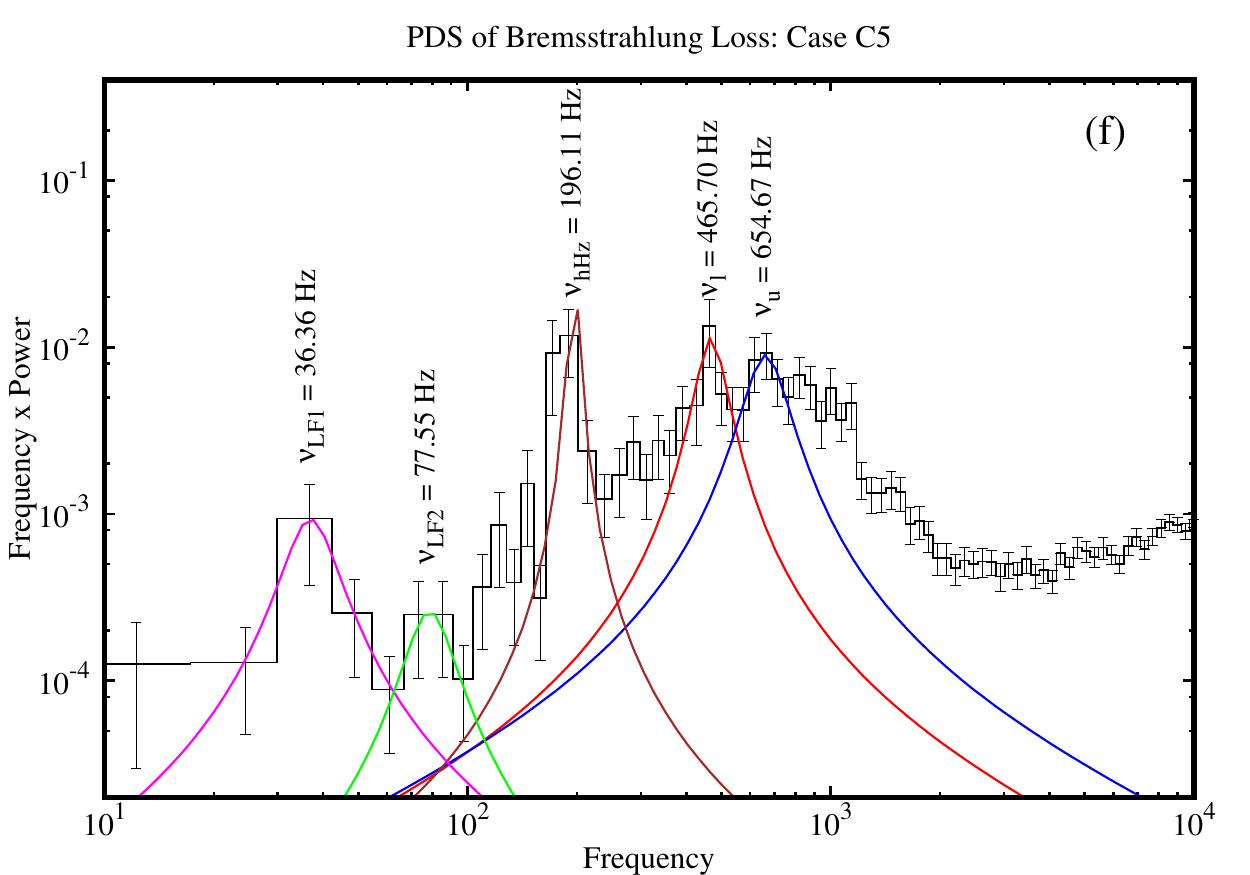}
\caption{(a)Power Density Spectrum ($frequency \times Power$ vs $frequency$) of the mass outflow from upper quadrant of the simulation setup, for case $C1$. A clear QPO at $\sim 127~Hz$ with a harmonic at $\sim 254~Hz$ is observed. (b) A similar QPO feature at $\sim 129~Hz$ was found in the corresponding lightcurve of the bremsstrahlung loss. (c) PDS of bremsstrahlung loss for $C2$, (d) for $C3$, (e) for $C4$ are also shown. The clear harmonics of the $kHz$ QPOs and the influences of those in the power spectra have been ignored while fitting separate peaks (b-e). In (f), for case $C5$, only the 2nd harmonic of $\nu_{LF}$ is shown. For that, five distinct peaks have been fitted with Lorentzian profiles, having centroid frequencies at 36.36 Hz (magenta, $\nu_{LF1}$), 77.55 Hz (green, $\nu_{LF2}$), 196.11 Hz (brown, $\nu_{hHz}$), 465.70 Hz (red, $\nu_{l}$), and 654.67 Hz (blue, $\nu_{u}$) which are also plotted.}
\label{fig1}
\end{figure*}

\section{Conclusions}

In this paper, we have made an effort to understand the dynamics of an inviscid, rotating, geometrically thick and optically thin flow around a weakly magnetic neutron star using smoothed particle hydrodynamics. We add modified bremsstrahlung cooling so as to particularly study 
the timing properties of the centrifugally driven shocks (CENBOL) as well as the density jump (normal boundary layer or NBOL) formed on the star surface due to 
sudden arrest of infalling matter. Such simulations were done {\it for black holes} earlier (MLC94, MSC96, MRC96, M01a, M01b, ACM02, CAM04, GC13, GGC14, Deb et al. 2017) and oscillations of the CENBOL were found in radial and vertical directions. These oscillations were then identified with the low frequency QPOs observed in black hole candidates.

In presence of a hard boundary on the neutron star surface,
we expect another oscillation of higher frequency as well as others due to non-linear interactions of the 
flows. Our present simulations indeed show complex timing properties of the radiation as well as the flow dynamics. For this, we chose the flow to have an accretion rate ($\dot{m}_h$) below the Eddington limit for all the cases we studied. The exponent $\alpha$ in the cooling rate was varied from $0.5$ (bremsstrahlung) to $0.6$ to observe the 
effects of the strength of cooling on the oscillation of shocks. The specific angular momentum $\lambda$ was chosen based on the recent study by Deb et al. (2017) where an onset of vertical oscillation was seen between the two values chosen here. The variation of the radius of the neutron star had a more complex effect as it controlled the effects of radiation pressure on the hydrodynamics through the parameter $\mathcal{C}$ when a self-consistent variation is chosen. It is also to be noted that for the timescales of simulations C1 to C5, the mass and momentum deposited on the star were negligible as compared to the star's mass and spin. This accumulation of mass and momentum would become significant for timescales of the order of years or more, which is outside the purview of the current paper. However, the energy release at the surface was found to be significant, resulting in a measurable change of $T_{NS}$.

We show, among other things, that the simulations produce both low and high-frequency QPOs and the oscillations last during the whole simulation period (more than 200 dynamical timescales measured at the injected flow radius, i.e., $30~r_S$). This suggests that the QPOs are formed due to a part of the flow dynamics
and not a transient effect as inferred by others (e.g., Barret and Olive, 2005). 
We measure the QPO frequencies and find that both the centroid frequencies and Q factors match well with observed results of neutron stars such as GX17+2, 4U 1728-34 and Cir X-1. We believe that the advective flow suggested in the literature while explaining the behavior of the source Cir X-1 (Boutloukos et al. 2006), may be the same as the
dynamic transonic flow solution we discuss here. We showed that the presence of angular momentum itself can generate multiple modes of oscillation in CENBOL and NBOL, manifesting as QPOs in the PDS, in presence of cooling. 
In Fig. 1(b), 2(c-d), 3(a), we see different types of shocks are being formed. The outer shock was found to be vertical near the equatorial plane and oblique away from the plane, very similar to
what was seen for simulations around black holes (MLC94). The bending instabilities reported in M01a, are also found here and correspond to the hecto-Hz oscillations found in the PDS. 

So far, in the literature, a model which appears to be capable of phenomenologically addressing both timing and spectral properties is the transition layer (TL) model of TLM98 who assumed the disc to be Keplerian to begin with. The viscosity was also assumed to be high enough to maintain a Keplerian distribution. The QPOs are then explained as the oscillation of the TL at different orbital frequencies. An extended TL was used in the COMPTT and COMPTB models for the analysis of spectra of accreting NSs. Many LMXBs have been studied using the COMPTB framework, such as 4U 1728-34 (Seifina et al. 2011), GX 3+1 (Seifina and Titarchuk 2012), GX 339+0 (Seifina et al. 2013), 4U 1820-30 (Titarchuk et al. 2013), Scorpius X-1 (Titarchuk et al. 2014), 4U 1705-44 (Seifina et al. 2015) etc. The HMXB 4U 1700-37 has also been examined using the same model (Seifina et al. 2016). Two COMPTB components were needed in general for spectral fitting. The one corresponding to a cloud closer to the star had a relatively lower temperature and the one closer to the Keplerian disc typically had higher temperature (cited works above). The one corresponding to Comptonization of NS surface photons, showed a saturation in COMPTB model's spectral index (Farinelli and Titarchuk, 2011). This index is different from the spectral index found by fitting the power-law component of the spectrum. The latter can have a continuous range of values depending on the two accretion rates as shown in BC17. However, since the source of high viscosity required to sustain a complete  Keplerian distribution remains elusive and physical processes to create two Compton clouds out of a Keplerian disc is also not demonstrated, we preferred to start with a sub-Keplerian inviscid advective flow onto a neutron star which is a general configuration.  In presence of higher turbulent viscosities, this flow will become a Keplerian disc easily as in the case of black hole accretion by simply redistributing angular momentum (Chakrabarti, 1990, 1996, Chakrabarti 2017), and thus in softer states, the flow could resemble a configuration similar to the TLM98 model. In general, the flow should have both the sub-Keplerian and Keplerian components (Chakrabarti, 1995; 1997, 2017, CT95). The infall timescales from Chakrabarti 1995 (see also, Chakrabarti 1997) have the similar $r^{3/2}$ dependence with the radial distance $r$. The absolute value only differs from Keplerian orbital time scale by a factor of $R_{comp}/2\pi$, $R_{comp}$ being the shock compression ratio. This, in principle, suggests that the same numerical values of frequencies for all such sources could be produced by the TCAF scenario, even when the flow is not-Keplerian to begin with and the viscosity is too low to form any TL. The pre-shock and post-shock regions of NBOL are orders of magnitude denser than those of CENBOL and contribute to higher frequency oscillations as well. In the TCAF scenario, the physical NS boundary aids in the formation of inner shock (NBOL) apart from the the CENBOL surface, which is formed even in BH accretion.

We found that the ratio $\nu_{u}/\nu_{l}$ varied from $1.38$ to $1.53$ and the ratio $\nu_{l}/\nu_{hHz}$ varied from $1.66$ to $4.20$ in our simulations. The Q factors of lower kHz QPOs were always found to be greater than the Q factor of the upper one. This agrees well with observational results (Barret et al. 2005). In our cases, the separation $\delta \nu$ between $\nu_{u}$ and $\nu_{l}$ increased with accretion rate, which is similar to results of Cir X-1 (Boutloukos et al. 2006). In fact, the values of Table 1, when scaled with the mass of Cir X-1, lies in the same ball park figure as that found in observation. 

In passing, we may mention that our simulation clearly demonstrated that the 
advective flows onto a non-magnetic neutron star create a stable configuration. The flow with a sub-Keplerian specific angular momentum has at least two density jumps in accretion and shocks were also found in the outflows. In presence of cooling, the shocks underwent oscillations in both radial and vertical directions and were manifested as QPOs in the PDS of bremsstrahlung loss from the system. We also find that the shock location, and hence the QPO frequencies depend on many parameters of the flow, viz., the specific angular momentum $\lambda$, the accretion rate $\dot{m}_h$, the radius of the star $R_{NS}$ and the strength of cooling $\alpha$, which we studied here. 

The most general flow configuration should depend on the spin, mass and radius of the star and the viscosity of the flow. In this paper, we kept the spin frequency to be constant at 142 Hz, the mass was kept to be constant at 1.0 and viscous effects were ignored. In a future paper we will vary these crucial parameters and general a spectrum of steady and 
oscillating solutions. Non-local cooling processes as well as electron energy distribution due to shock acceleration
are yet to be incorporated in the simulation. From the results of BC17, it is clear that when accretion rates are high for a hot accretion flow, Compton cooling has a significant effect on the temperature of the flow. In future, we wish to couple Monte Carlo simulation with this hydrodynamic code to get time-dependant spectral variation. The results would be 
presented elsewhere.

\section*{Acknowledgement}
AB would like to acknowledge Prof. D. Molteni and Dr. G. Lanzafame for their 2D SPH simulation source code, written by them with SKC, 
which was modified for case of neutron star with cooling and particle coalescing schemes to produce the results used in this paper.

\end{document}